\documentclass[12pt,preprint]{aastex}
\def\Msun{{\,M_\odot}}
\def\Zsun{{\,Z_\odot}}
\slugcomment{{\it Accepted by the Astrophysical Journal}}

\begin{document}
\shorttitle{The Colors of dE GCS, Nuclei, and Stellar Halos}

\title{The Colors of Dwarf Elliptical Galaxy Globular Cluster Systems, 
Nuclei and Stellar Halos}

\author{Jennifer M. Lotz \altaffilmark{1}, Bryan W. Miller \altaffilmark{2},
and Henry C. Ferguson \altaffilmark{3}}

\altaffiltext{1}{Department of Astronomy and Astrophysics, University of
California, Santa Cruz, CA 95064; jlotz@scipp.ucsc.edu}
\altaffiltext{2}{Gemini Observatory, Casilla 603, La Serena, Chile;
 bmiller@gemini.edu}
\altaffiltext{3}{Space Telescope Science Institute, Baltimore, MD 21218; 
ferguson@stsci.edu}

\begin{abstract}
We present the results of a Hubble Space Telescope WFPC2 F555W and F814W 
survey of 69 dwarf elliptical galaxies (dEs) in the Virgo and Fornax 
Clusters and Leo Group. The $V-I$ colors of the dE globular clusters, 
nuclei, and underlying field star populations are used to trace the dE 
star-formation histories.  We find that the dE globular cluster candidates
are as blue as the metal-poor globular clusters of the Milky Way.
The observed correlation of the dE globular cluster systems' $V-I$ color 
with the luminosity of the host dE is strong evidence that the globular 
clusters were formed within the the halos of dEs and do not have a 
pre-galactic origin. Assuming the majority of dE clusters are old, the 
mean globular cluster color- host galaxy luminosity
correlation implies a cluster metallicity $-$ galaxy luminosity relation of 
$\langle Z_{GC}\rangle \propto L_B^{0.22 \pm 0.05}$, which is significantly 
shallower than
the field star metallicity - host galaxy luminosity relationship observed 
in Local Group dwarfs ($\langle Z_{FS} \rangle \propto L^{0.4}$).
The dE stellar envelopes are $0.1-0.2$ magnitudes redder in $V-I$ than 
their globular clusters
and nuclei.  This color offset implies separate star-formation episodes
within the dEs for the clusters and field stars, while the very blue 
colors of two dE nuclei trace a
third star-formation event in those dEs less than a Gyr ago.  
\end{abstract}

\keywords{galaxies: dwarfs ---galaxies: nuclei --- galaxies: star clusters 
--- galaxies: stellar content --- galaxies: evolution --- 
clusters: individual(Virgo, Fornax, Leo)}

\section{INTRODUCTION}
It is generally believed that the dwarf galaxies we observe in the 
local universe are survivors of the 
hierarchical growth of massive galaxies.  While present-day dwarf 
galaxies may somehow be different from the majority of proto-galactic 
fragments which coalesced into larger objects, the formation of stars 
in the smallest survivors are an important test of our understanding
of galaxy formation and evolution.  If star-formation quickly follows 
the gravitational collapse of the initial density perturbations in the 
early universe, we might expect
the stars in dwarf galaxies to be quite old because their halos should be among
the first objects to form.  Dwarfs in large-scale over-densities destined
to become clusters and groups should collapse and form stars at earlier 
times than galaxies formed in under-dense regions (Tully et al. 2002).   
Yet star-formation must not proceed too
quickly, otherwise the baryons in the universe would become 
locked up in stellar remnants, 
leaving no fuel for the star-formation we see locally (e.g. 
Kauffmann, White, \& Guiderdoni 1993).

Dwarf elliptical galaxies (and their low-mass Local Group analogues, 
dwarf spheroidals or dSphs) are an obvious place to look for ancient stars and to 
track early star-formation.   dSphs and dEs in the Local Group are almost 
completely free of gas and typically
dominated by stars older than several billion years. These objects range 
from dEs with $M_{B} \geq -18$ and total masses $\leq 10^{10} \Msun$ down 
to dSphs with $M_{B} > -11$ and total masses  $\geq 10^6 - 10^7 \Msun$.  
While the smallest dSphs have masses and luminosities comparable
to massive star clusters, they are physically different entities.  
Dwarf galaxies sit inside the gravitational potential wells of dark 
matter halos, and the kinematics of their stars are increasingly dominated
by this underlying dark matter halo (e.g. Aaronson 1983;
but see Geha, Guthathakurta, \& van der Marel 2002), whereas star clusters 
do not have associated dark matter halos (Pryor et al. 1989; Moore 1996).  
Massive globular clusters are compact, high-surface brightness objects, 
while dE/dSphs are diffuse and extended with decreasing surface-brightnesses 
with decreasing luminosities (Kormendy 1985).
Most star clusters consist of a single population of stars with uniform 
age and chemical composition. On the other hand, studies of the 
color-magnitude diagrams and spectroscopic abundances of individual 
stars in Local Group dEs/dSphs show surprisingly complex star-formation 
and chemical enrichment histories in even the faintest galaxies 
(e.g. Carigi, Hernandez, \& Gilmore 2002; Grebel 1997; 
Smecker-Hane et al. 1994).

The complex star-formation histories of the nearest dEs/dSphs are 
probably regulated by a number of mechanisms.  Because of their low 
masses and shallow gravitational potential wells, dwarf galaxies are 
potentially fragile systems. The radiation, stellar winds, 
and supernovae produced by the first generations of stars can 
photo-dissociate, shock-heat, and expel much of the remaining gas, 
thereby suppressing future star-formation.  Dekel and Silk (1986) and 
later work by Dekel \& Woo (2003) showed that the onset of 
supernovae-driven winds could explain the observed scaling relations of
metallicity and mass-to-light ratio with total luminosity for galaxies 
less massive than $\sim 10^{10} \Msun$.  External heating by the high 
background of ultraviolet photons prevalent 
at early times in the universe (the epoch of re-ionization) can 
photo-ionize and even photo-evaporate small gaseous halos and 
prevent star-formation altogether (Efstathiou 1992; Babul \& Rees 1992; 
Bullock, Kravtsov, \& Weinberg 2000).  Dwarf halos with virial 
velocities $< 15$ km s$^{-1}$ which collapse after re-ionization may 
never accrete gas or form stars and remain dark (Bullock et al. 2000), 
while more massive halos and those which collapse prior to re-ionization 
in over-dense regions may produce some stars, albeit very inefficiently 
(Dong, Murray, \& Lin 2003).  Dwarf halos may also be rejuvenated by 
accreting new gas and/or cooling sufficiently to form stars after the 
UV-background has sufficiently cooled 
at $z \sim 1$ (Babul \& Rees 1992, Efstathiou 1992).   Finally, 
star-formation in dwarf galaxies may be strongly dependent on 
their environments and interactions with other galaxies.  
dEs are found exclusively as the satellites of massive galaxies or in 
rich clusters of galaxies.
The rich cluster environment can ram-pressure strip accreted field 
dwarfs of their gas (Lin \& Faber 1983)
and transform or tidally destroy dwarfs via multiple encounters with 
more massive galaxies (Moore, Lake, \& Katz 1998; Kravtsov, Gnedin, 
\& Klypin 2004).

Disentangling the effects of these different processes on the 
evolution of dEs is challenging.
In the Local Group, we have the advantage of proximity and can 
resolve the stars in $\sim 20$ dEs/dSphs.
The color-magnitude diagrams and spectroscopic abundances of 
these stars can then be used to
trace the star-formation and chemical enrichment histories of 
the dSph (eg. Grebel 1997). 
In the Virgo and Fornax Clusters are thousands of dEs, but 
because of their distances, current (largely 
ground-based) studies of these objects have been limited to 
the integrated colors and spectra. 
In this study of Leo, Virgo and Fornax Cluster dEs, we still cannot 
examine the properties
of their individual stars.  However, with HST resolution, we can 
distinguish their bright star clusters 
and nuclei from the stellar envelopes of field stars and use the 
colors of these three sub-populations to
constrain the evolution of cluster dEs.

dEs and dSphs possess increasing globular cluster specific 
frequencies (number of globular clusters 
per host galaxy luminosity normalized to $M_V = -15$) at 
fainter luminosities (Miller et al. 1998).
Studies of the globular cluster systems of Local Group 
dSph and a few dEs outside of the Local Group 
indicate that the majority of dE globulars are as old and 
metal-poor as the oldest 
globular clusters in the Milky Way ([Fe/H]$ < -1.0$, $>$10~Gyr), 
and are significantly
more metal-poor than their associated field star populations 
(Strader et al. 2003;  Puzia et al. 2000; 
Buonanno et al. 1999; Da Costa \& Mould 1998; Durrell et al. 1996a,b).  
A handful of intermediate-age 
and metal-rich clusters are also found in the Local Group dSph 
(van den Bergh 2000).
High-resolution numerical simulations of the collapse of dwarf-sized 
$10^8 \Msun$ dark matter halos 
predict the formation of dense $10^5 \Msun$ star clusters within 
dwarf halos at redshifts 
greater than 10 (Bromm \& Clarke 2002).
However, the origin of metal-poor globular clusters as ``pre-galactic'' 
(Peebles \& Dicke 1968;
Burgarella et al. 2001;  Beasley et al. 2003) or associated with galaxy 
formation and assembly 
(Searle \& Zinn 1978; Cote et al. 1998; Bromm \& Clarke 2002; 
Kravtsov \& Gnedin 2003) remains controversial.   

Bright dEs often have luminous compact stellar nuclei. 
The nature and formation of the dE nuclei are poorly understood. 
The fraction of nucleated Virgo dEs increases with both dE luminosity 
(Binggeli \& Cameron 1991) and position within the cluster 
(Ferguson \& Sandage 1989).  
In dense environments such as the centers of clusters of galaxies, the 
pressure from the surrounding inter-galactic medium may allow dwarf galaxies
to retain their gas during star formation and produce multiple generations 
of stars (Babul \& Rees 1992), forming nuclei in the process.
Spectroscopy of several bright Fornax dE nuclei show Balmer lines stronger than
the metal-rich globular cluster 47 Tuc (Held \& Mould 1994;
Brodie \& Huchra 1990), and NGC 205's nucleus is very blue and has 
strong Balmer line absorption 
(Jones et al. 1996; Ho, Filippenko, \& Sargent 1995). These examples suggest that 
the stars in many dE nuclei may have formed quite recently.  However, Sagittarius's 
probable nucleus is the 
massive, old, and metal-poor globular cluster M54 (Sarajedini \& Layden 1995).
The expected dynamical friction timescales for 
massive globular clusters in the low-density dEs is quite short,  
therefore many nuclei could simply be 
bright clusters that have spiraled into the center (Lotz et al. 2001; 
Oh \& Lin 2000; Hernandez
\& Gilmore 1998), and contain predominately old and metal-poor stars.

The integrated colors and spectral features of the field stars in 
nearby cluster dEs are 
consistent with the intermediate and old mean ages ($\sim 4-13$ Gyr) and 
intermediate 
metallicities ($-1.0 < $ [Fe/H] $<$ 0) derived for most Local Group dE/dSphs 
(e.g. Rakos et al. 2001;  Conselice, Wyse, \& Gallagher 2003; 
Geha et al. 2003).
The details of the past star-formation events are impossible to 
recover from the 
integrated colors and spectral features of a complex stellar population; 
however spectroscopic measurements of the abundance of alpha elements 
(produced 
in both Type I and II SN) as compared to iron (primarily produced in 
Type I SN) 
give $\alpha$/Fe ratios equal to the solar value and below (Thomas et al. 2003; 
Geha et al. 2003).  
This implies that the majority of stars in dEs were produced more than 1 
Gyr after initial 
onset of star-formation and after Type I SN chemically-enriched the dE 
interstellar medium. 

In this paper, we examine the $V-I$ colors of the globular clusters, 
nuclei, and field stars of 
69 dwarf elliptical galaxies observed with the Hubble Space Telescope.  
In \S 2, we describe our globular cluster detection and photometry, and 
the photometry for the dE 
stellar envelopes. In \S 3, we present the observed colors for each of 
these sub-populations and 
their dependence on the host galaxy properties and environment.  In \S 4, 
we discuss the implications of
the observed correlations for the formation of dE globular cluster systems 
and the star-formation
histories of dEs.  

\section{HST dE SNAPSHOT SURVEYS I, II, \& III}
The HST Dwarf Elliptical Galaxy Snapshot Survey has observed 69 dEs 
(Table \ref{de_tab}) in
the Leo Group and the Virgo and Fornax Clusters with the Hubble Space 
Telescope Wide Field/Planetary Camera 2 
(HST WFPC2, GO Programs 6352, 7377, and 8500).
The dE sample spans 6 magnitudes in $M_B$ ($-12 < M_B < -18$)
and has 45 nucleated and 24 non-nucleated dEs. 
HST WFPC2 images in filters F555W (2 $\times$ 230 s) and F814W (300 s) were
taken of each galaxy, with the galaxy centered on chip WF3. 
The data were corrected for instrumental signatures by the standard
HST calibration pipeline. 
A combined, cosmic-ray rejected F555W image was created with the 
cosmic-ray rejection routine COSMICRAYS (R. White, priv. comm.) and 
used as a template to identify and remove cosmic-rays from the single F814W image. 

\subsection{Globular cluster candidate detection and photometry}
Globular clusters are compact bright star clusters, with typical absolute
 magnitudes
$-4 > M_V > -11$ and half-light radii $\sim 3-4$~pc (Ashman \& Zepf 1998). 
The high angular resolution of the WFPC2 images ($\sim 0.1\arcsec$ per pixel) 
allows us to distinguish bright globular cluster candidates and 
nuclei from the underlying diffuse halo of dE field stars.
At the distance of the Leo Group and the Virgo and Fornax Clusters, 
globular~clusters
are not truly resolved in short exposures by the WF camera and may be 
treated as point sources.  
The IRAF DAOPHOT point-source detection algorithm was run on the
combined, cosmic-ray corrected F555W image to identify compact
sources with fluxes $3 \sigma$ above the background within  a 0.2~\arcsec\ 
radius aperture. 
The F555W and F814W fluxes for all detected point sources were measured within
this aperture, and corrected for the missing flux in the wings of the PSF 
(i.e. aperture-corrected). 
The average aperture corrections for the WF chips are $-0.275 \pm 0.014$ 
magnitudes for F555W and $-0.307 \pm 0.015$ magnitudes for F814W
(see \S 2.4 in Miller et al. 1997 for description of how these
corrections are calculated).  
We also corrected the flux of the globular cluster candidates for the 
charge transfer
efficiency effects across the WF chips (Whitmore \& Heyer 1995). 
We adopted the Holtzman et al. (1995) photometric calibrations 
 to transform our F555W and F814W photometry to the standard Johnson $V$ 
and Cousins $I$ values.
Foreground Galactic extinction and reddening values were taken from 
Schlegel et al. (1998, courtesy of the NASA Extragalactic Database) for 
each galaxy.
Because most dEs are gas-poor systems,
we assumed no internal extinction. Only FCC46 
shows evidence for internal dust and active star-formation 
(Drinkwater et al. 2000).
We adopted distance moduli
of 30.0 for the Leo Group, 30.92 for the Virgo Cluster, and
31.39 for the Fornax Cluster, based on the Cepheid distances calculated 
by the HST
Key Project (Freedman et al. 2001).  It is important to note that these 
are average distances
to these clusters and that the Virgo and Fornax Clusters have a depth 
along the line of sight of 
$\sim 1-2$ Mpc (Neilsen \& Tsventanov 2000). We do not have independent 
distance estimates for each dE, 
and so the absolute magnitudes  for the globular clusters and dEs have
distant-dependent uncertainties on the order of $\pm 0.15$  magnitudes.

\begin{figure}
\epsscale{1.0}
\plotone{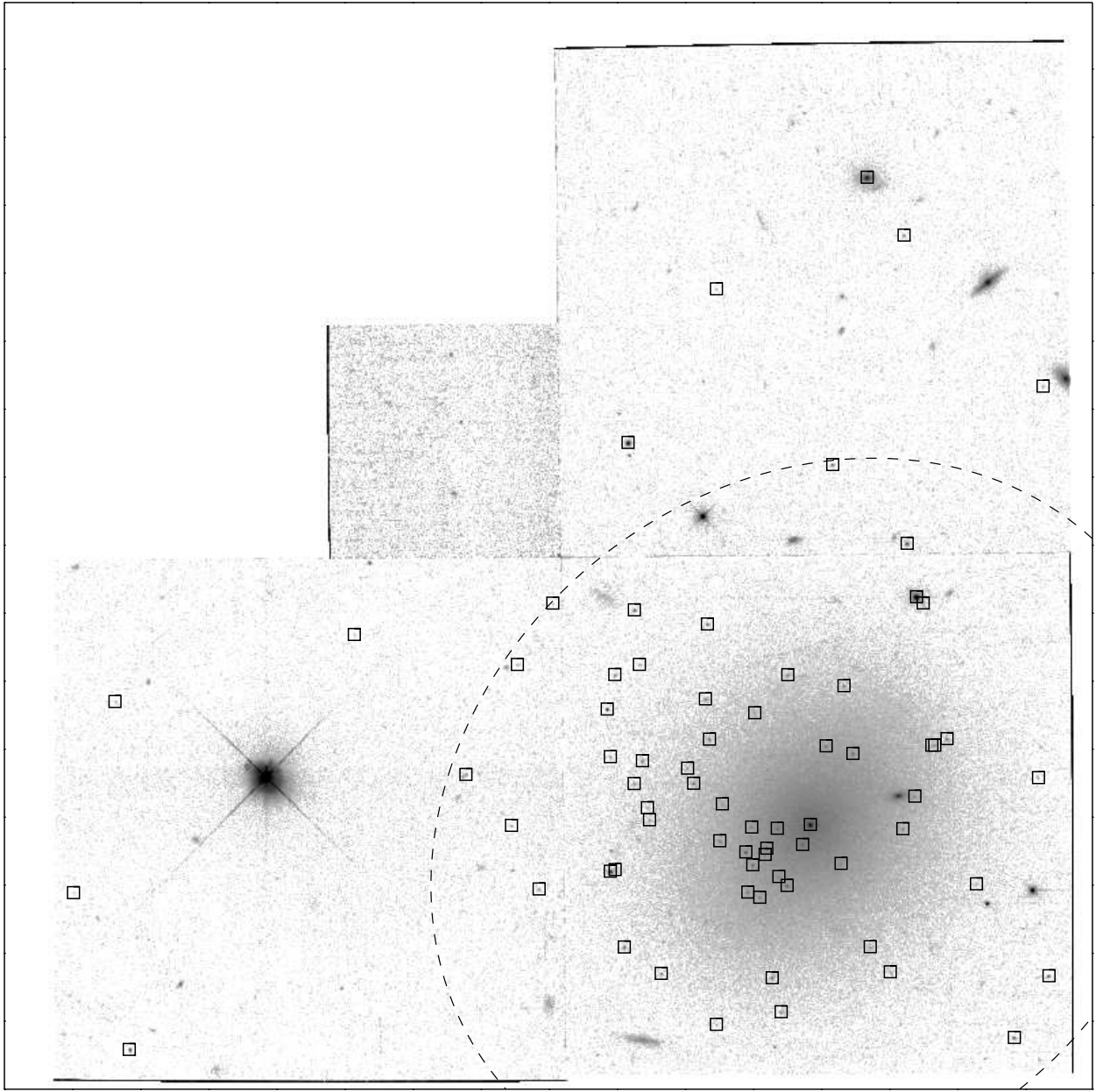}
\caption[WFPC2 F555W mosaic image of nucleated dE VCC940]
{The WFPC2 F555W mosaic image of nucleated dE1 VCC940 ($M_B = -16.1$).  Compact sources
meeting our globular cluster candidate criteria are boxed.  Objects which 
fall on WF3 or within 5 scalelengths of the center of the galaxy (dashed ellipse)
are globular cluster candidates.  Objects outside of this radius are treated as
background objects. \label{vcc940}}
\end{figure}

Globular cluster candidates (GCCs) were selected based on size (FWHM$_{F555W}$ 
$<$ 2.5 pixels), $V-I$ color ($0.5 < V-I < 1.5$ and error $(V-I)$ $< 0.3$), 
and proximity to the galaxy
(see Miller et al. 1997 for discussion of globular cluster candidate
 color and size selection).   Compact background galaxies and foreground stars
within our color criteria may be mistaken for globular cluster candidates, 
and there
may exist intergalactic globular clusters which are not bound to the dEs.
We must correct for this contamination when estimating the number of globular
clusters associated with each dE and examining their colors.   In Tables 
\ref{nde_tab}
and \ref{nonde_tab}, we have assumed that objects meeting our candidate 
criteria
on WF chips 2 and 4 and outside of an elliptical aperture centered on the 
dE with semi-major
axis length equal to 5 times the dE's exponential scalelength are background 
contaminants
(Figure \ref{vcc940}).  The number density per area of these background 
objects were
used to correct the number of observed GCCs on WF chip 3 and those on  WF 
chips 2 and 4 
that lie within 5 scalelengths of each dE.

Monte Carlo simulations of our point source detection indicate that 
our detections become incomplete at $V \sim 25.7$.  Assuming that dE 
globular clusters 
have a intrinsic luminosity function similar
to the Milky Way globular cluster luminosity function, we detect 
$\sim 85\%$ of the globular clusters in the Fornax dE images, 
90\% of the globular clusters in the Virgo dE images,  
and 98\% of the globular clusters in the Leo dE images.  

\subsection{Galaxy photometry}
The HST snapshot observations are not well suited to measuring the total 
galaxy 
luminosities.  The flux of brightest dEs spills out onto chips WF2 and WF4, 
whereas  
the faintest (and lowest surface brightness) dEs are barely detectable above 
the sky noise.  
Stiavelli et al. (2001) fit the surface brightness profiles for a subset of 
25 dEs 
(observed during cycle 6) with Sersic profiles and derived total magnitudes 
from 
their fits.  They also measured the color profiles and find little
evidence for radial color gradients. Therefore the color measured within a 
fixed aperture 
should be a reliable estimate of the galaxy's true color.

We measured the relative fluxes of $V$ and $I$ within a fixed isophotal 
aperture for 45 dEs in 
our sample with central surface brightnesses $\mu_V < 24$ magnitudes per 
\sq \arcsec, using the
galaxy photometry package SExtractor (Bertin \& Arnouts 1996).   
In order to improve the signal to noise for the galaxy detection algorithm, 
the F555W and F814W images 
were summed together and used as the SExtractor detection image. The 
isophotal aperture for each galaxy was set at 1$\sigma$ above the sky level 
in the 
summed image; a lower detection threshold resulted in numerous spurious 
noise detections around the 
perimeter of the galaxy and a higher detection threshold failed to find 
the dEs.  
The galaxy flux within this aperture was then measured for the F555W 
and F814W images separately. SExtractor also corrects the measured galaxy 
fluxes for extended 
background galaxies observed directly behind the galaxy. 
The flux from the 
compact globular cluster candidates and the central nuclei was included in 
the isophotal flux of the galaxy.   
Most nuclei are too faint to contribute significantly to the measured
colors of the dE stellar envelopes. The two brightest nuclei are the 
red nucleus of VCC1254 ($M_V = -12.25, V-I = 1.01$) and the very blue
nucleus of FCC46 ($M_V = -12.16, V-I = 0.58$).  For these two galaxies,
we masked out the nuclei and re-measured the stellar envelope color.
We found no change in the color of FCC46, while VCC1254 became 0.05 magnitudes redder.
Given that the errors on the galaxy colors are $\geq 0.03$, we expect that the 
rest of the nuclei will have a negligible effect on the dE halo colors.

The measured galaxy colors were transformed to the Johnson-Cousins
photometric system using the same equations as the globular clusters 
(Holtzman et al. 1995).  
Due to the extended nature of the galaxies, no charge transfer efficiency 
corrections 
were applied to the galaxy colors.  Ignoring this correction adds an 
additional $\leq$ 1.5\% uncertainty. 
The galaxy colors were de-reddened using the Schlegel et al. (1998) 
foreground extinction values.   The galaxies' surface brightness profiles 
were measured from 
the WFCP2 F555W images in elliptical apertures using the IRAF task ELLIPSE 
and fit to exponential profiles:
\begin{equation}
I(r) = I_0 \ {\rm exp}(-r /r_0)
\end{equation}
where $I_0$ is the central intensity and $r_0$ is the exponential scalelength.

\section{COLORS OF dE STELLAR POPULATIONS}

\subsection{dE Globular Cluster Systems}
 Although dEs have large numbers of globular clusters for their 
luminosities (Miller et al. 1998), they are faint and possess
at most a handful of clusters.  We generally find less than 20 GCCs
above the background counts for each dE (Tables \ref{nde_tab} and 
\ref{nonde_tab}).  
Because of the relatively high fraction of
background contamination ($\sim 30-50\%$),  determining the mean color 
and the color distribution for
each dE globular cluster system is impossible without spectroscopic 
confirmation of the
globular cluster candidates.  
Instead, we have constructed {\it composite} cluster color distributions
for all the dEs within a given absolute magnitude range. By binning the globular 
cluster 
candidates by host galaxy luminosity, we improve our number statistics, 
can correct for the background contamination, 
and look for broad trends with dE luminosity. 

\begin{figure}
\epsscale{0.75}
\plotone{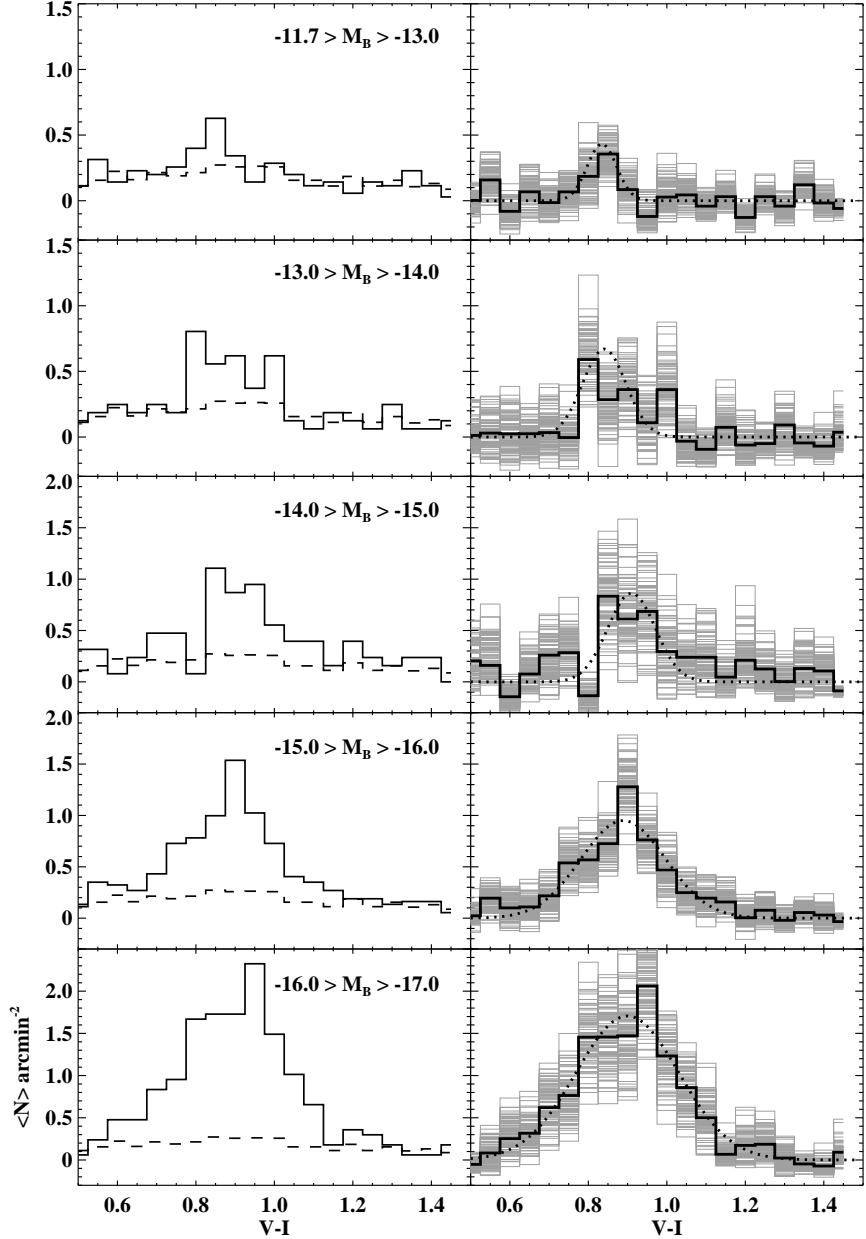}
\caption
{The composite globular cluster candidate color distribution as a function of host dE luminosity. Left: The
raw galaxy GC (solid line) and background (dashed line) distributions. Right: The background 
corrected GC color distributions (heavy solid line), the first 100 bootstrapped GC color distributions
(gray lines), and the median Gaussian fit to $10^4$ bootstrapped color distributions (dotted curve).\label{vihist}}
\end{figure}

A $V-I$ color histogram
is created for all of the GCCs for the dEs within a luminosity range and 
normalized to the
total area of the galaxies sampled (left of Figure \ref{vihist}).  Then 
the $V-I$ color histogram for all background objects (also
normalized to the total area of the background detections) is used to 
correct for the background object color
distribution.  The resulting background-corrected GCC color histograms 
are shown in heavy solid lines in the 
right of Figure \ref{vihist}.  Also shown in Figure \ref{vihist} are the 
color histograms
produced by ``bootstrapping'' or re-sampling at random the original sample of 
background objects and dE GCCs 100 times (gray histograms) to illustrate 
the uncertainty
of each derived color distribution.  Each bootstrapped distribution 
contains the same number of background objects and dE GCCs as the original 
sample, but 
does not require that each replacement data point be unique.

Most of the dE GCCs are as blue in $V-I$ 
as the metal-poor globular clusters in the Milky Way and other bright galaxies
(Harris 1991;  Forbes \& Forte 2000; Kundu \& Whitmore 2001).  
We detect very few globular cluster candidates
bluer than $0.6-0.7$. This is also consistent with the minimum globular 
cluster 
metallicity observed in the Milky Way ([Fe/H] $\sim -2.5$, Laird et al. 1988). 
Assuming the dE globular cluster candidates are as old as typical Milky Way 
globular clusters ($\geq$  10~Gyr), their color range 
agree with the metallicity range found for the globular cluster observed
in Local Group dwarf spheroidal galaxies (dSphs)
and nearby dEs ( $-2.5 <$ [Fe/H]$ < -1.0$; see Table \ref{lg_gc_tab} ). 

However, the dE GCC color distributions are not uniform.  
We find that the GCC color distribution significantly broadens 
and the mean $V-I$ color becomes slightly redder with increasing
dE luminosity (Figure \ref{vihist}). We have randomly re-sampled the 
original background
objects and dE GCCs 10,000 times and fit single Gaussians to the binned and 
background-corrected color distributions with the IDL routine GAUSSFIT 
in order to derive peak $V-I$ colors and $\sigma (V-I)$.  In Figure 
\ref{vihist},
we have plotted the Gaussian fits with the median peak color for the 
10,000 bootstrapped
color distributions (see also Table \ref{vi_tab}).  This median color 
distribution has 
a peak $V-I = 0.83 \pm 0.02$ and $\sigma (V-I) = 0.03 \pm 0.01$ for the 
faintest dEs ($-11.7 < M_B < -13$)
and a peak $V-I = 0.90 \pm 0.01$ and $\sigma (V-I) = 0.13 \pm 0.01$ for 
the brightest dEs
($-16 < M_B < -17$).   We find that a linear fit to the observed peak 
GCC color - dE luminosity trend gives
a shallow but significant slope :
\begin{equation}
\langle V-I \rangle_{GC} = - 0.018(\pm 0.006)\times {\rm M_B(dE)}\ +\ 
0.606(\pm 0.093)
\end{equation}
with $\bar{\chi}^2 = 0.46$.  

\begin{figure}
\epsscale{1.0}
\plotone{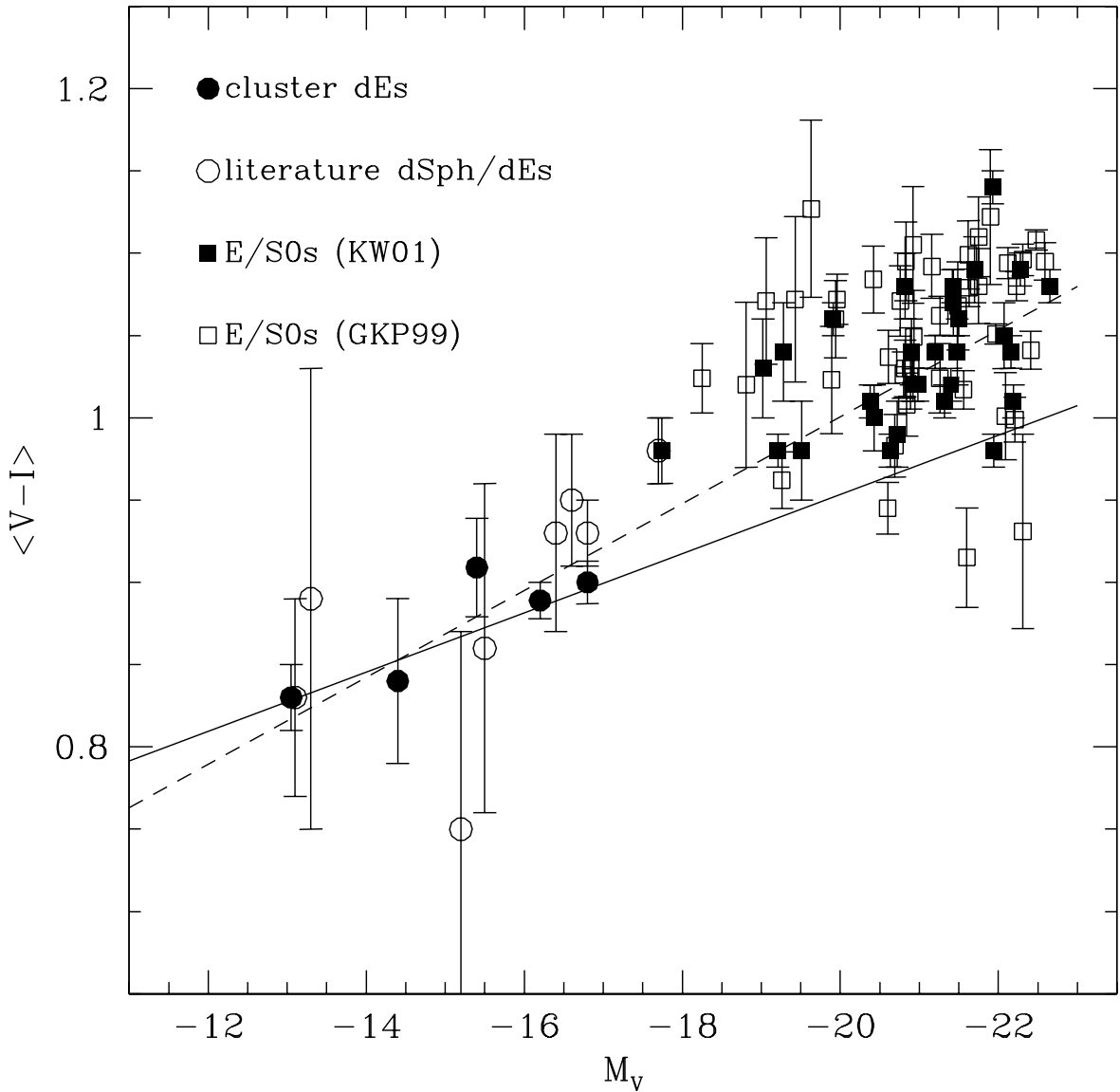}
\caption
{The  peak globular cluster $V-I$ color v. host galaxy luminosity for our sample of Leo, Virgo and
Fornax cluster dE GCS as well as nearby dE/dSphs, E and S0 GCS drawn from the literature (Table 5,
KW01, GKP99).  The error-bars here reflect the uncertainty
in the peak of the color distribution.  The lines are the linear fits to
 the dE $\langle V-I \rangle_{GC} - M_V$ correlation (solid: our dEs only; dashed: 
our dEs plus literature dEs/dSphs).\label{viall}}
\end{figure}

In Figure \ref{viall}, we compare the  peak $V-I$ colors to the 
average colors of the literature dEs (Table \ref{lg_gc_tab}), giant 
elliptical galaxy (E) and S0 
globular cluster systems (Kundu \& Whitmore 2001, Gebhardt \& Kissler-Patig 
1999; here after KW01 and GKP99).  
We have converted the literature values for the mean [Fe/H] of the globular 
cluster systems 
of five Local Group dE/dSphs, as well as three dEs outside of the 
Local Group, to a $V-I$ color  with the empirical globular cluster $V-I$ - 
metallicity calibration of Kissler-Patig et al. (1998):
\begin{equation}
{\rm [Fe/H]} = -4.50 + 3.27(V-I)
\end{equation}
(This conversion implicitly assumes that the globular clusters are 
relatively old.)
The E/S0 globular cluster systems show a great deal of scatter in their mean 
colors, and the E/S0 sample spans a small range in luminosity, thus 
there has been some debate in the literature over the existence of a GC color - 
host galaxy luminosity trend (Harris 1991; Ashman \& Zepf 1998; 
Strader, Brodie \& Forbes 2004).
However, for all but a few galaxies, the average colors of the E/S0 colors 
are consistent with 
the color-luminosity trend observed in dE sample, as are the 8 nearby 
dE/dSph GCS.  
Including the $V-I$ colors derived from the literature mean [Fe/H] values for
dSph and dE, we find
\begin{equation}
\langle V-I \rangle_{GC} = -0.026(\pm 0.005)\times {\rm M_B(dE/dSph)}\ +\ 
0.491(\pm 0.079)
\end{equation}
with $\bar{\chi}^2 = 0.93$.
Including the 70 E/S0 $\langle V-I \rangle_{GC}$ colors from GKP99 and KW01 
in our linear fit
gives a worse $\bar{\chi}^2 = 3.31$ but a similar slope for the GC color - 
$V$-band
luminosities:
\begin{equation}
\langle V-I \rangle_{GC} = -0.026(\pm 0.001)\times {\rm M_V(dE/dSph/E/S0)} 
+\ 0.484(\pm 0.024).
\end{equation}

\begin{figure}
\epsscale{1.0}
\plotone{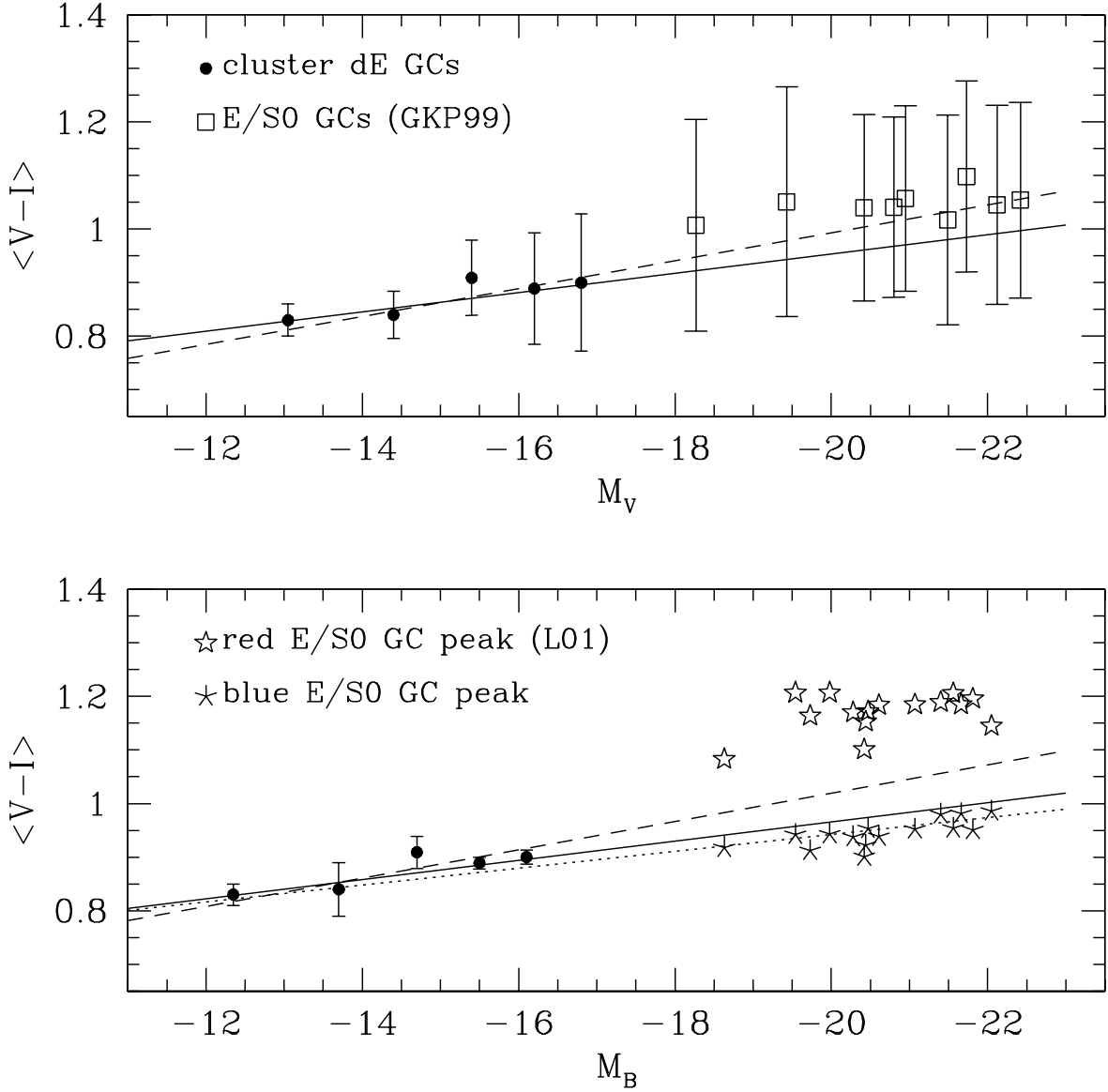}
\caption
{The dispersion in the globular cluster color distribution. {\it Top}: The peak $V-I$ values for the
entire GCS v. host galaxy $M_V$. The error bars give the dispersion about this peak 
$\sigma (V-I)$. 
{\it Bottom:} The peak $V-I$ colors of the dE GCS and the red and blue peak colors of 16 E/S0s (Larson et al. 2001)
v. host galaxy $M_B$.  The solid and dashed lines are the linear fits to 
dE $\langle V-I \rangle_{GC} - M_V$ correlation, 
and the dotted line is the linear fit to blue E/S0 GCs from Larson et al.(2001). Here the
error bars are standard errors of the mean.
\label{visig} }
\end{figure}

The globular cluster color distributions are clearly broader for  E/S0s 
than the dEs.
In Figure \ref{visig}, we illustrate the increasing breadth of the color 
distributions as a  function of host galaxy luminosity.  
In both panels of Figure \ref{visig}, 
the dE GCC peak colors are plotted as a function of dE $M_V$ (assuming 
$M_V = M_B - 0.7$ where
$M_V$ is unavailable) and
the Gaussian color dispersions $\sigma(V-I)$ given in Table \ref{vi_tab} 
are shown as the error-bars.  We have excluded the literature
dE/dSph GCS from these plots because of the small numbers of GCs
in each system and the uncertainties associated with calculating a
robust mean color and dispersion for very small samples (see Strader
et al. 2004).
In the top panel, we compare our dEs to the fitted $V-I$ color distributions for
43 E/S0 globular cluster systems taken from GKP99.  
They fit the $V-I$ color distributions with a bi-weight estimate of
the location and scale, which correspond to a Gaussian mean
and $\sigma$ to first order.  We have binned the E/S0 GCS by host galaxy 
luminosity, 
and plot the mean GKP99 peak $V-I$ values, with error-bars to
indicate the mean scale (roughly equivalent to $\sigma (V-I)$), along side 
our dE globular cluster peak 
colors and $\sigma(V-I)$ (Figure \ref{visig}) as a function of host galaxy 
luminosity.  
While the overwhelming majority of E/S0 globular cluster systems have 
$\sigma (V-I)$ larger than 
those observed in the brightest dE, there is no correlation between 
$\sigma (V-I)$ and E/S0 
luminosity. 

In the bottom panel of Figure \ref{visig}, we plot the dE 
$\langle V-I \rangle_{GC}$ and $\sigma (V-I)$ with the results of 
bi-modal Gaussian fits to ``red'' and ``blue'' globular cluster peaks 
for 16 E/S0s (Larsen et al. 2001).  
The fitted peak colors for the E/S0 blue globular clusters 
are consistent with the extrapolation of the peak GC color- host galaxy 
luminosity trend
observed in our sample of dEs (Equation 2), but are significantly bluer than the 
correlation derived when the literature dE/dSph cluster systems are included 
(Equation 4). 
Strader et al. (2004) also converted the metallicities of 9 dSph/dI 
globular clusters systems 
to $V-I$ color using a different relation and found that their colors 
were consistent with 
the color-luminosity dependence observed the blue GCs of ellipticals 
and spirals. 
While all galaxies possess globular clusters as blue as those in dEs, 
the brighter galaxies also 
possess increasing proportions of globular clusters redder than 
$V-I \sim 1.1$.  
The broadening in color distribution combined with the increasing number
of the reddest globular clusters is what gives rise to the trend observed 
in Figure \ref{vihist}, rather 
than a shifting narrow distribution.  

\subsection{dE field stars}
Early work on the Fornax dE population by Caldwell \& Bothun (1987) 
found that dEs followed 
the $U-V$ vs. $B$ color-magnitude relation observed for the Fornax giant 
ellipticals. 
However, dEs showed increased scatter in their colors and nucleated dEs 
were redder 
than non-nucleated dEs at the same luminosity.
The observed color-magnitude slope was consistent with the 
metallicity-luminosity relation found for Local
Group dSph ($Z \propto L^{0.4}$, Dekel \& Silk 1986).  Similar E/dE 
color-magnitude slopes
have been subsequently observed in the Fornax (Cellone et al. 1994, 
Rakos \& Schombert 2004), Coma (Secker 
\& Harris 1996), 
and Perseus Clusters (Conselice et al. 2002), again with increased scatter 
towards the
fainter luminosities. Spectroscopic abundances have confirmed that cluster 
dEs and Local Group dSphs 
follow the metallicity-mass correlation of giant ellipticals 
(Geha et al. 2003; Thomas et al. 2003). 
However, several more recent studies of the Fornax Cluster dE/dSph populations
which probed luminosities as faint as $M_V \sim -10$ have derived 
slightly shallower $V-I$ color 
v. $V$ magnitude slopes that imply $Z \propto L^{0.3}$ 
(Hilker, Mieske, \& Infante 2003; Karrick,
Drinkwater, \& Gregg 2003).  

\begin{figure}
\plotone{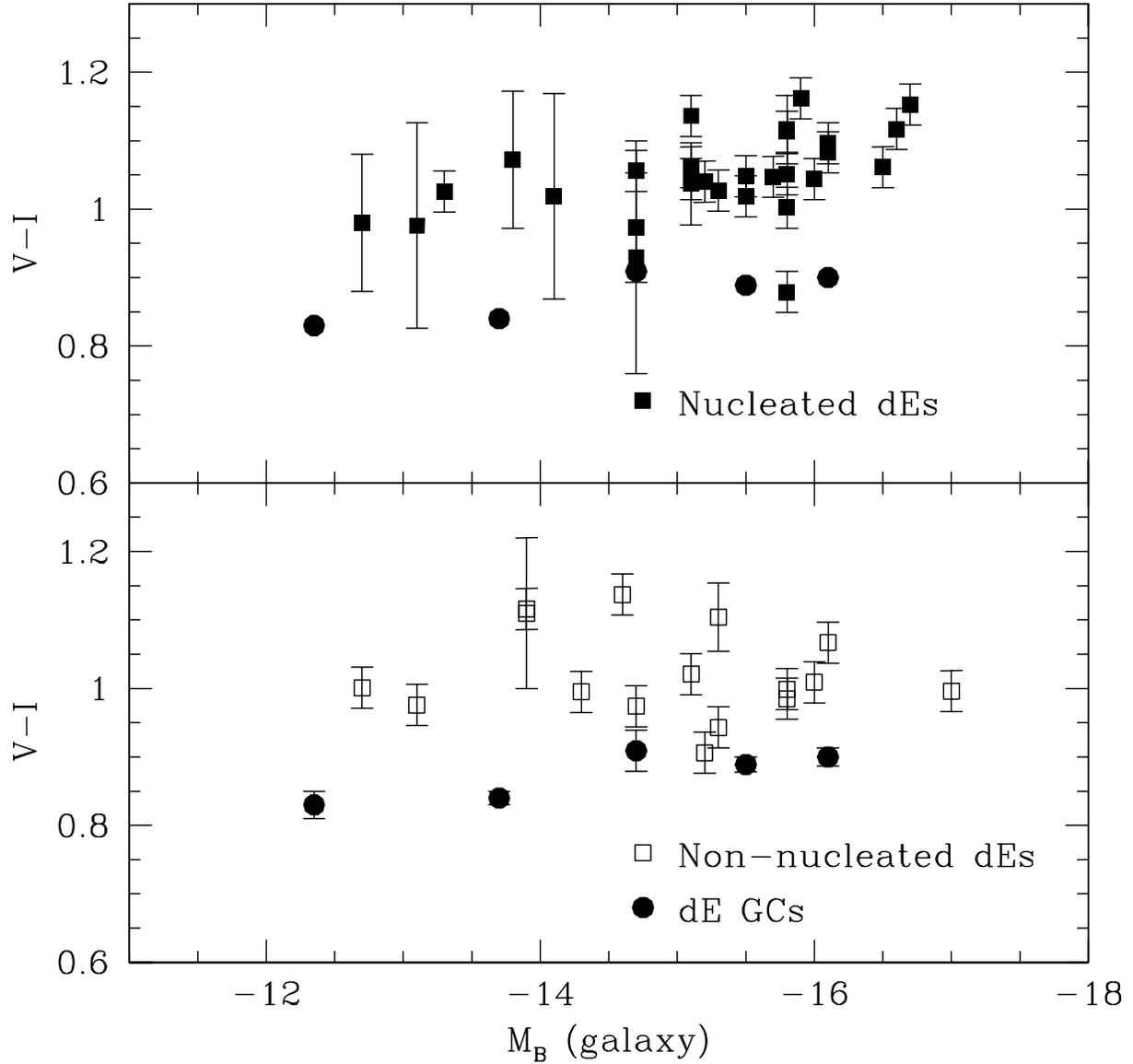}
\caption[Nucleated and non-nucleated dE colors]
{Nucleated (top) and non-nucleated (bottom) dE stellar halo colors v. host galaxy luminosity. 
The filled circles are the dE globular cluster colors from Figure \ref{viall}. \label{vigal}}
\end{figure}

We find that the integrated dE field star populations are redder than
their globular clusters by $0.1-0.2$ magnitudes (Figure \ref{vigal}).  
When we examine the nucleated dE and non-nucleated dEs separately, we
find evidence for a color-luminosity trend for the nucleated dEs, whereas
no such trend is evident for the non-nucleated dEs.  A Spearman rank 
correlation 
test finds a 99.7 \% probability of a color-luminosity correlation for 
nucleated dE, 
and  a 44.8 \% probability of correlation for non-nucleated dE.  
The entire dE sample
gives a 94.6 \% probability of a color-luminosity correlation.  
The derived color-magnitude relations are:
\begin{equation}
\begin{array}{ccl}
(V-I)_{FS} & = -0.031 (\pm 0.010) \times M_B ({\rm dE}) + 0.577 
(\pm 0.153) & {\rm for\ dE,N\ only} \\
(V-I)_{FS} & = -0.014 (\pm 0.005) \times M_B ({\rm dE}) + 0.829 (\pm 0.009) 
& {\rm for\ all\ dE} 
\end{array}
\end{equation}
The slope of the color-magnitude relation for the dE,N
matches the shallower trends found by Hilker et al. (2003) and Karrick et al. 
(2003), 
but is not inconsistent with the slightly steeper slope for cluster 
ellipticals.  
When both nucleated and non-nucleated dEs are considered,
the color-magnitude relation has a slope more like than found for 
the globular cluster systems  but with a higher zeropoint.

\begin{figure}
\plotone{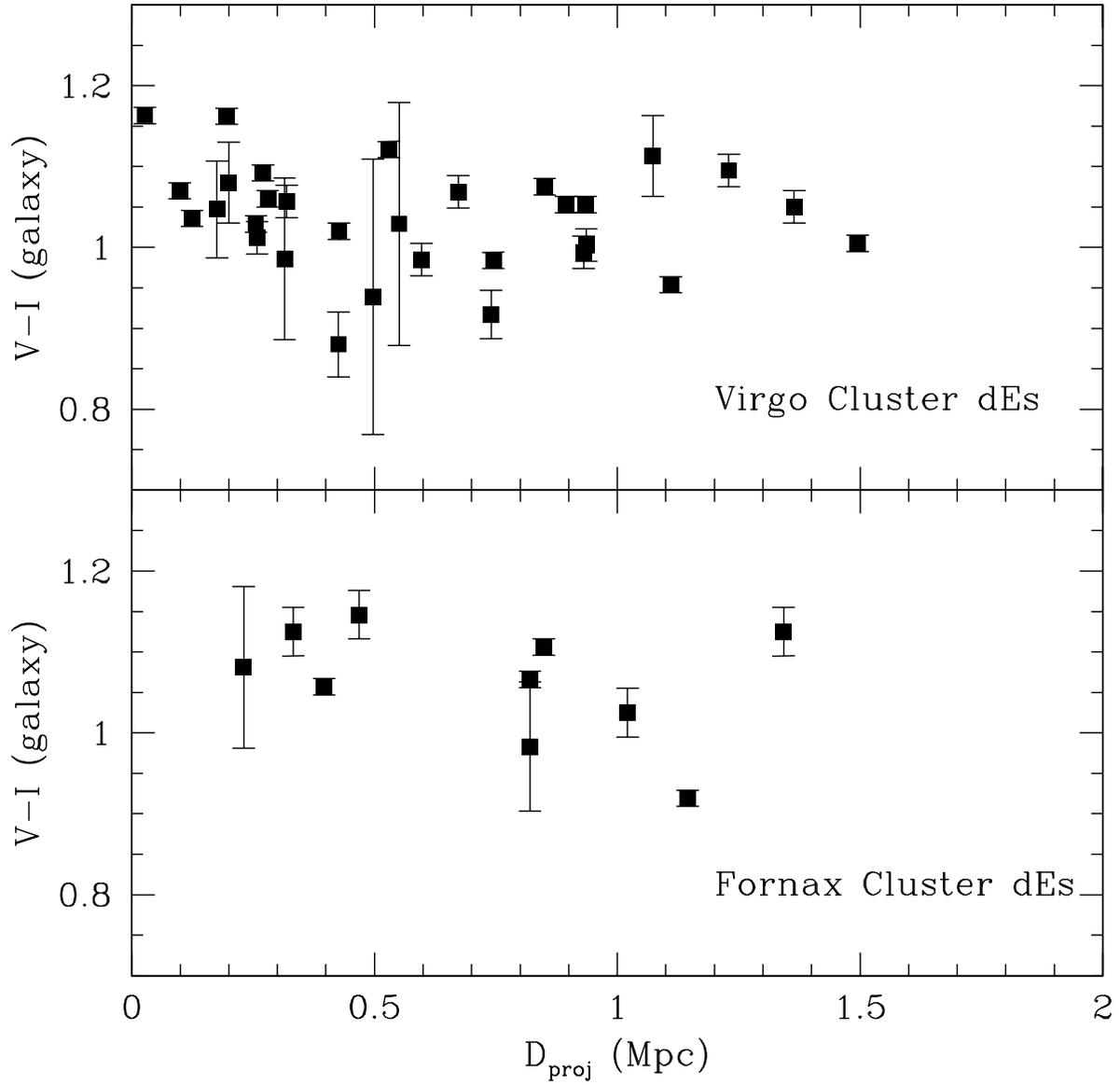}
\caption[dE halo colors v. projected spatial position]
{dE $V-I$ halo colors v. projected radius (Mpc) for Virgo and Fornax Cluster dEs.\label{vigaldist}}
\end{figure}

Several groups have claimed evidence for gradients in the color of dE
population with distance from the cluster centers (Secker 1996; 
Rakos et al. 2001), 
which would imply age or metallicity dependence on cluster position. 
In Figure \ref{vigaldist},  we plot galaxy color as a function of 
projected distance from the (closest) central cluster galaxy (M51 for the Leo
Group, NGC1399 for Fornax, and either M87 or NGC4472 for Virgo).
We find no obvious correlation between dE stellar halo color and the 
projected radial distance
from the central cluster galaxy for both the Virgo and Fornax cluster dEs 
($p < 59$\%). 

\subsection{dE nuclei}
We identify the nucleus of each dE,N galaxy as a bright compact object within 
1.5~\arcsec\ the dE's isophotal center.  
All nuclei meet our cluster candidate criterion (implying sizes $< 10$~pc), and 
are measured via the aperture photometry described in \S 2.2.  
We find 6 galaxies with  previously unidentified
nuclei (FCC46, FCC150, FCC242, FCC246, VCC646, and VCC1577).  VCC9 was 
originally classified as nucleated by Binggeli et al. (1985), 
but its brightest globular cluster candidate is 1.8~\arcsec\ from its center.  

\begin{figure}
\plotone{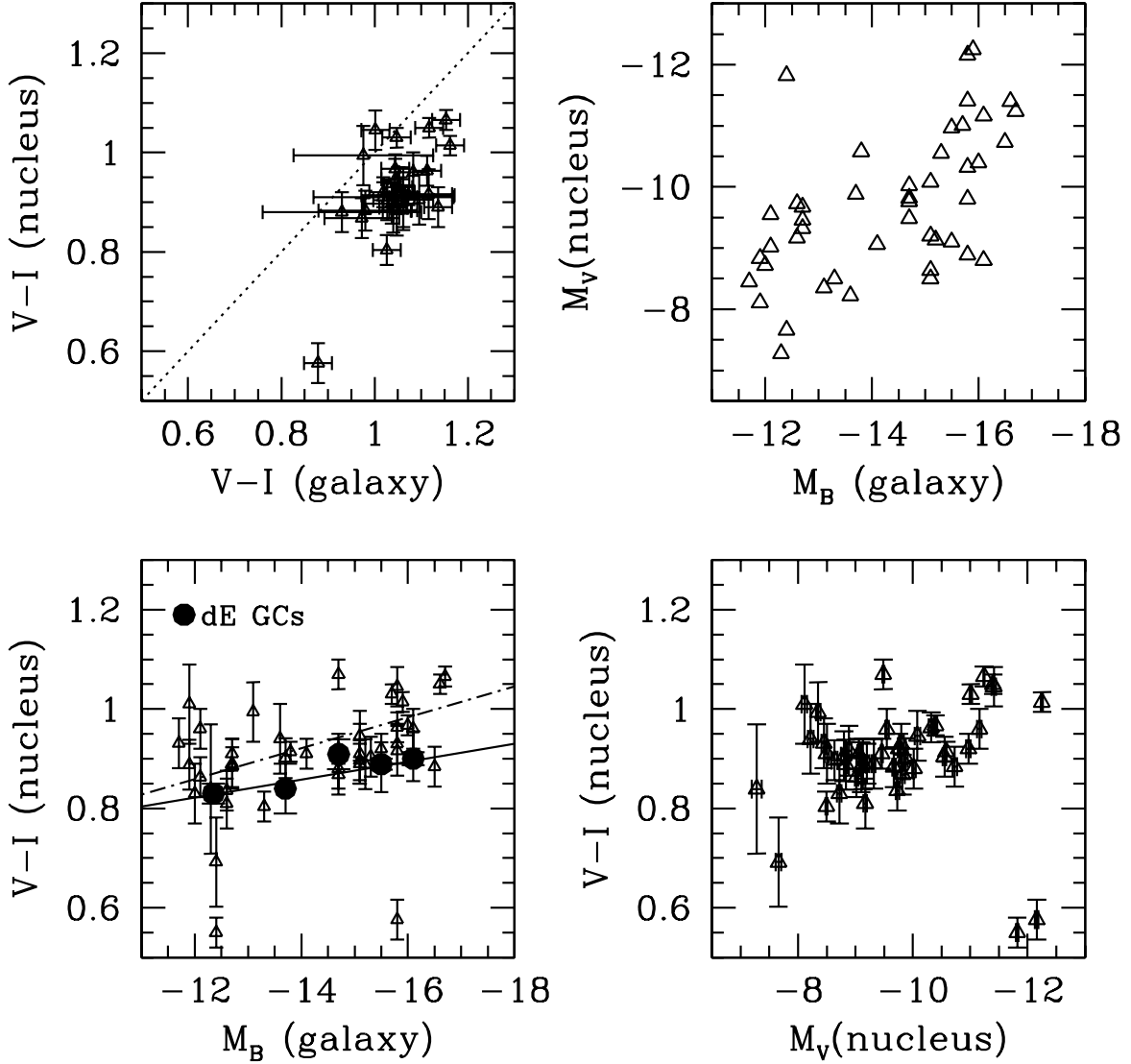}
\caption
{The dE nuclear properties.  {\it Top left:} 
The nuclear $V-I$ color vs. host galaxy  $V-I$ color. 
{\it Top right:} dE nuclear $M_V$ vs. host galaxy $M_B$. The outlier at
$M_V$ (nucleus) $= -12$, M$_B$(galaxy) $= -12.5$ is the very blue nucleus of VCC1714.
{\it Bottom left:} The dE nuclear colors (triangles) and peak globular cluster colors (circles) 
vs. host galaxy luminosity. The solid line is the fit to the dE GCs, and the dot-dashed line
is the linear fit to the dE nuclei excluding the two blue outliers.
{\it Bottom right:} Nuclear $V-I$ colors vs. nuclear $M_V$.  \label{nuc}}
\end{figure}

The dE nuclei are significantly bluer than their stellar 
envelopes (top left of Figure \ref{nuc}), contradicting previous 
ground-based studies (Caldwell \& Bothun 1987; Rakos \& Schombert 2004).
Two dEs possess very blue nuclei with $V-I < 0.6$ (VCC1714 and FCC46). 
We find a correlation between nuclear color and dE color ($p = 98.9$\%),
where redder nuclei are found in redder and brighter dEs 
and one of the bluest nuclei lies in a very blue dE (FCC46).
Nuclear luminosity is a strong function of the host galaxy's luminosity, 
although with a $\sim 1$ magnitude dispersion not accounted for by the 
photometric uncertainties (top right of Figure \ref{nuc}).

The nuclei colors are also correlated with host galaxy luminosity at the
99.5\% confidence level (bottom left of Figure \ref{nuc}), with a
linear fit giving
\begin{equation} 
(V-I)_{nucleus} = -0.031(\pm 0.004) \times M_B({\rm dE}) + 0.487(\pm 0.055)
\end{equation}
When the two very blue nuclei are excluded, nuclear color is also found 
to be correlated with nuclear luminosity at the 99.5\% level (bottom right 
of Figure \ref{nuc}), with a slope as shallow as for the nuclear color- host 
galaxy luminosity trend (bottom left of Figure \ref{nuc}).

The nuclei are often slightly redder than the 
dE globular cluster systems (bottom left of Figure \ref{nuc}).
Many nuclei are much brighter than typical globular cluster
candidates, which have a peak in their luminosity function at 
$M_V = -7.4$ (Miller 2003).  Because nuclei become redder as they increase in luminosity, 
we have compared the faint ($M_V \leq -10$) nuclei colors to those
of the dE globular cluster candidates with the same range of
absolute magnitudes. We find no color difference between the
dE nuclei and globular clusters at fixed nucleus/cluster luminosity, 
suggesting that the faint nuclei are indistinguishable from 
bright globular clusters. 

\begin{figure}
\plotone{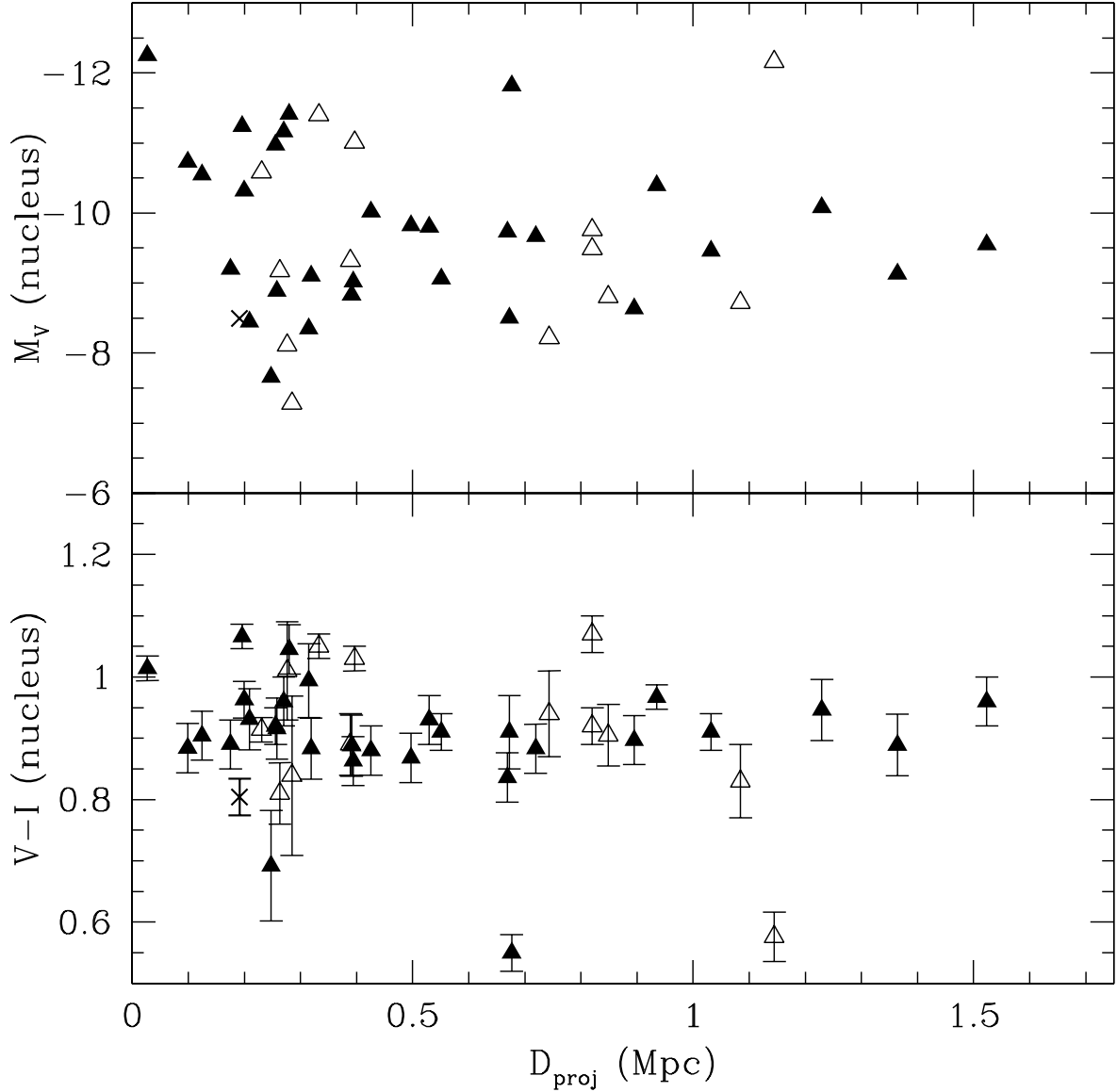}
\caption[Nuclear properties v. host dE's projected spatial position]
{Nuclear properties v. host dE's projected distance from the cluster center. 
(filled triangles: Virgo Cluster,  open triangles: Fornax Cluster,  crosses: Leo Group) \label{nuc_r} }
\end{figure}

We have also looked for trends in the dE nuclear properties with position
in the Virgo and Fornax Clusters. As with the dE stellar envelopes, 
we fail to find any significant nuclear color or luminosity trends 
with distance from central cluster galaxy,  even when the two very blue nuclei 
are excluded from the sample ($p < 49.2$\%; Figure \ref{nuc_r}). 

\section{DISCUSSION}
\subsection{Constraints on Ages and Metallicities}
\begin{figure}
\plotone{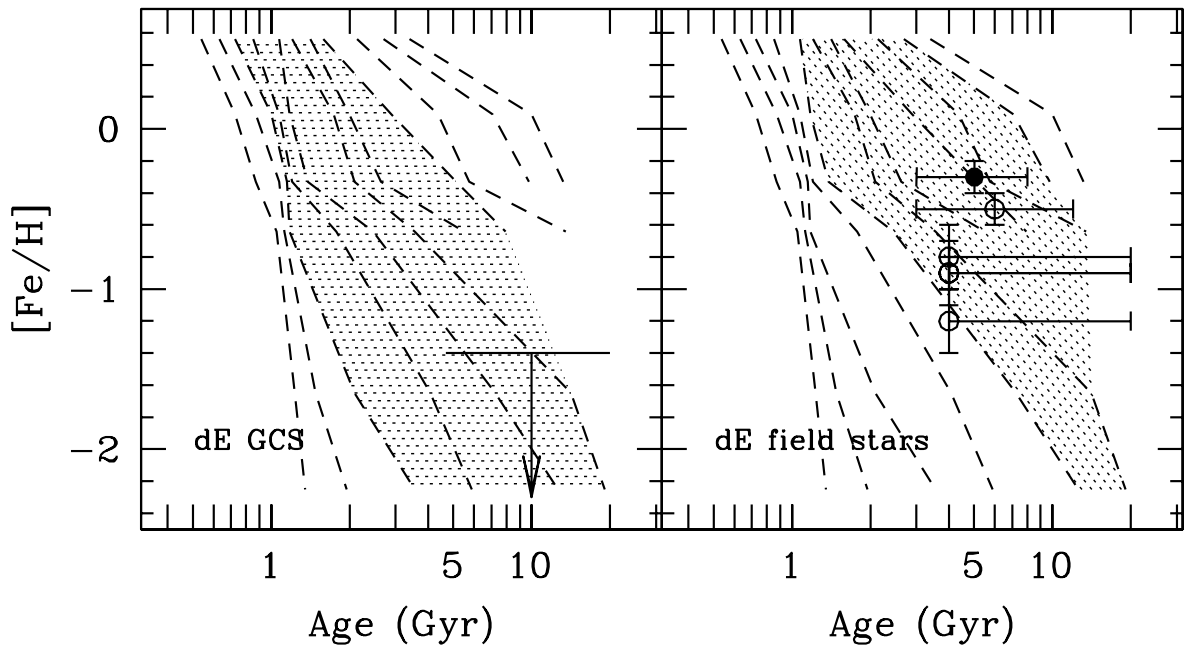}
\caption
{ The age-metallicity constraints on the dE stellar populations
from the $V-I$ colors and Bruzual \& Charlot (2003) single burst population synthesis models.
The dashed lines are lines of constant $V-I$ color ($V-I = 0.7 - 1.20$ in 0.05 intervals from
left to right). {\it Left:} the range of dE GCS colors (shaded region)
and the [Fe/H] and age constraints for Local Group dSph/dE old GCs (Table 5) are shown.  {\it Right:} 
the shaded region shows the range of dE stellar halo colors ($0.9 \le V-I \le 1.15$).  
The mean spectroscopic result from Geha et al. (2003)
for 17 Virgo and Fornax dEs is plotted as the solid circle. The constraints for the
field stars in the Local Group dSphs/dEs from Table 5 are plotted as open circles. \label{mods}} 
\end{figure}

The integrated $V-I$ color of a stellar population is primarily a function of the
temperature of the stellar population's red-giant branch, which becomes hotter and 
bluer with younger ages as well as lower metallicities.  Therefore 
with a single optical color, it is impossible to distinguish between
a young, metal-rich stellar population and an old, metal-poor stellar population. 
Nevertheless, we attempt to place some reasonable constraints on the relative
ages and metallicities for the dE stellar populations given the observed $V-I$ and
what we know about local dE/dSph systems.  Using the 
Bruzual \& Charlot (2003) population synthesis models for a single starburst with a
Salpeter stellar initial mass function and the Padova 1994 isochrones, 
we have plotted the range of possible ages and metallicities for the 
dE stellar halos ($0.9 < V-I < 1.15$) and dE globular clusters 
($0.8 < V-I < 1.0$) in Figure \ref{mods}.  In these plots, the dashed lines are lines 
of constant $V-I$ color, and the shaded regions show the range of observed colors and
possible ages and metallicities for each dE stellar population.
Most of the dE stellar halos have $V-I > 0.95$ (Figure 6), implying luminosity-weighted
ages $>$ 2 Gyr and [Fe/H]  $> -1.6$, consistent with the more precise ages and metallicities of dSph/dEs
derived via spectroscopic absorption lines and resolved stellar color magnitude diagrams
(Geha et al. 2003; see also Table 5).  The bluer
colors of the globular clusters (and nuclei) give much weaker constraints, but are
consistent with [Fe/H] $< -1.4$ observed for old dSph GCs (Table 5).
The Bruzual \& Charlot (2003) tracks for the colors of the GCs in the faintest dEs ($V-I < 0.85$) 
imply that these clusters are younger than 8 Gyr. (However, the older Bruzual \& Charlot 
models allow for ages greater than 8 Gyr if [Fe/H] $< -1.7$ and $0.8 < V-I <  0.85$.  Also,  
several Milky Way globular clusters have integrated $V-I < 0.85$, 
$-1.4 > $[Fe/H] $> -2.2$, and ages $\geq 10$ Gyr.)  The model tracks
also require that the two nuclei with $V-I < 0.6$ are younger than 1 Gyr.

The color offset between the dE stellar halos and their globular cluster systems
requires that the bulk of the dE field stars are either 
significantly more metal-rich or older than their globular clusters.  
If the dE globular cluster systems were bluer
because they were several Gyr younger than the dE field stars, one may expect them to be
$\sim 0.6$ magnitudes brighter than old Milky Way globular clusters of the same mass and
color.  However, initial estimates of the composite dE globular cluster 
luminosity function (GCLF) indicate that its characteristic luminosity is identical to 
that of the Milky Way ($M_V = -7.4 \pm 0.1$; Miller 2003).
Given that Local Group dE/dSph globular clusters systems are more metal-poor 
than their hosts' field stars and that the dE GCLF is indistinguishable from that of the Milky
Way, we will assume  that our sample of dE GCs is dominated
by old and metal-poor globular clusters for the rest of this discussion. 
This implies either separate star-formation epochs for the dE GCs and field stars, 
or a strong metallicity gradient for dE stars with the destruction of the inner and more 
metal-rich clusters and the inward migration of the outer, more metal-poor clusters.
Most dE nuclei span a similar color range as the dE GCs, and therefore could be coeval with the dE GCs.  
However, the existence of two very blue nuclei requires that some nuclei
are formed in a later star-formation event, after both the dE globular cluster and 
field star formation epochs. In this section, we discuss the possible formation histories 
of these three dE sub-populations 
(clusters, field stars, and nuclei) and the constraints on the evolution of the dEs. 

\subsection{dE GC formation}
The observed correlations between dE GC colors and host dE luminosities is strong
evidence that the GCs knew about the gravitational potentials of their host dEs during their formation and
were not accreted by the dEs at a later time.  This rules out a primordial or pre-galactic 
origin for the globular clusters in dEs (Peebles \& Dicke 1968). 
If the majority of the dE globular clusters are  old and roughly coeval, the observed
mean GCS color- host galaxy luminosity relation suggests a mean GCS
metallicity - host galaxy luminosity correlation of
\begin{equation}
\begin{array}{ccl}
\langle Z \rangle_{GCS} &\propto \ L_B^{0.15 \pm 0.05} & {\rm for\ our\ dEs\ only} \\
\langle Z \rangle_{GCS} &\propto \ L_B^{0.22 \pm 0.05} & {\rm for\ all\ dSphs\ \& dEs}\\
\langle Z \rangle_{GCS} &\propto \ L_V^{0.21 \pm 0.02} & {\rm for\ dSphs,\ dEs,\ \&\ E/S0s}
\end{array}
\end{equation}
assuming the $(V-I) -$ [Fe/H] calibration of Kissler-Patig et al. 1998.
The uncertainty in the color-metallicity calibration has been added to the errors
in quadrature.   If we assume the slighter steeper $(V-I) - $ [Fe/H] calibration for
the Galactic globular clusters from Barmby et al. (2000), we find that the exponents for 
derived GC metallicity - host galaxy relations are 0.04 - 0.06 greater, but agree with
those in Eqn. 8 to within the uncertainties. These correlations are significantly shallower 
than the metallicity-luminosity 
relation observed for the field star populations of local dwarf galaxies ($Z \propto L^{0.4}$, Dekel \& Silk 1986), 
and that predicted for GCS by recent SPH hierarchical galaxy evolution models ($Z_{GC} \propto
M_*^{0.5}$, Kravtsov \& Gnedin 2003).

\begin{figure}
\epsscale{0.8}
\plotone{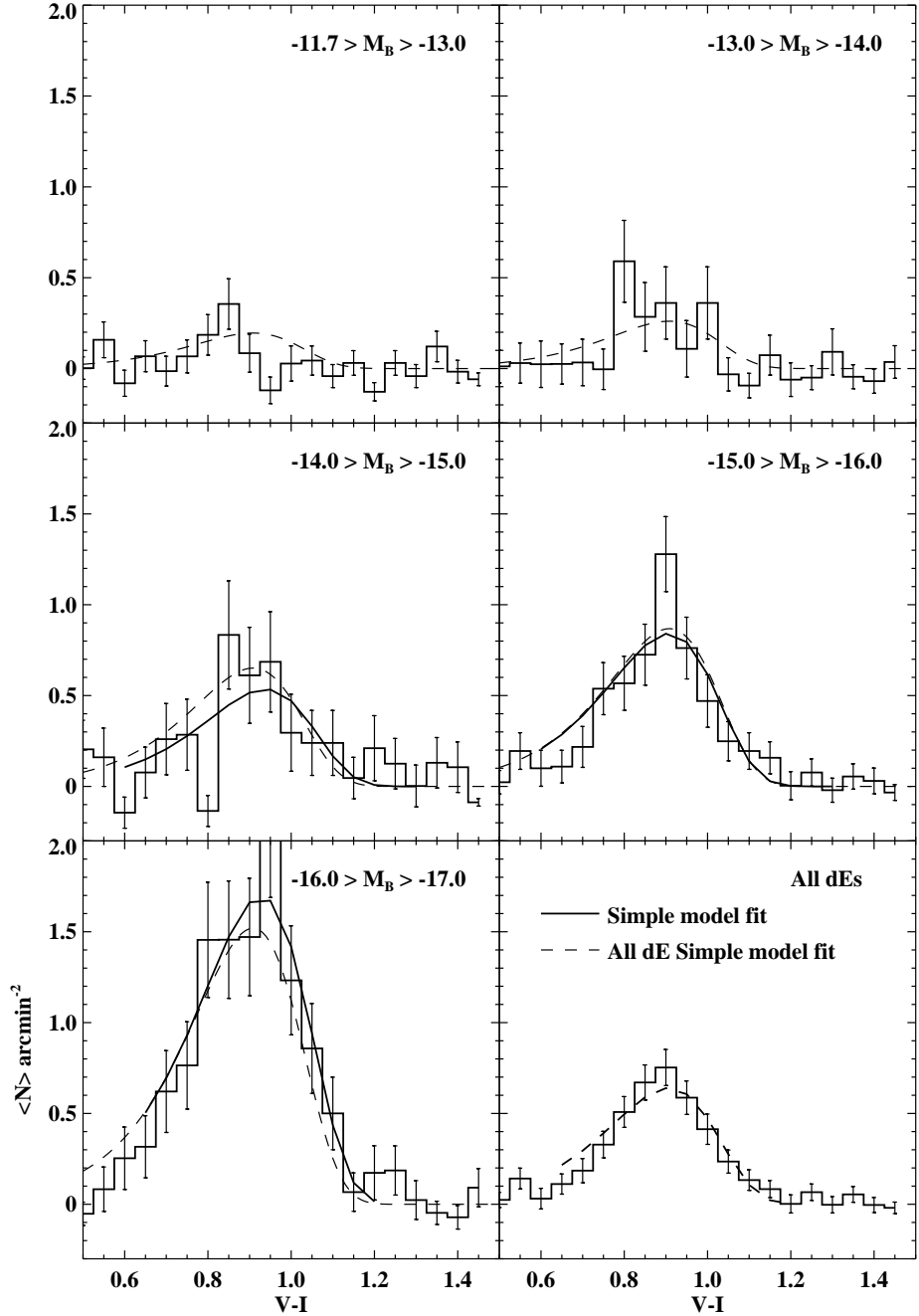}
\caption[``Simple model'' fits to dE globular cluster color distribution]
{Simple model fits to the dE GC color distributions, assuming Kissler-Patig et al. (1998)
$V-I$ to [Fe/H] calibration.  The solid curves are the best fits for each histogram,
and the dashed curves are from the best fit to all dE GCCs (lower right) for comparison. 
No reliable fits were found for the faintest GC color distributions (top panels).
\label{simple} }
\end{figure}

Assuming the dE GCs are older than 5 Gyr, 
the observed increasing color dispersion also implies an increasing dispersion in metallicity. 
The Simple model of galactic chemical evolution is the zeroth order model against
which to compare the  metallicity (and color) distribution of a stellar population (Tinsley 1975), or,
because  each globular cluster is a chemically homogeneous population, the metallicity distribution
of a system of globular clusters. The Simple model assumes that 1) the galaxy is a ``closed box'' with no 
inflow or outflow of gas; 2) the metals are well mixed; 3) the galaxy is initially pure gas with primordial abundances;
and 4) the nucleo-synthetic yields of stars are constant for ``primary elements'' (elements
created directly from hydrogen and helium). The metallicity distribution of stars $s$ for a Simple model
is
\begin{equation}
\frac{ds}{d\ {\rm log}(z)} \propto\  z\  {\rm exp}(-z) 
\end{equation}
where $z = Z/p$ and $p$ is the effective chemical yield.
In Figure \ref{simple}, we have fit Simple model metallicity distributions to the $V-I$ color
distributions, again assuming the Kissler-Patig et al. (1998) color-metallicity calibration. 
(Strictly speaking, the Simple model distribution applies only to primary
elements which follow the instantaneous recycling approximation, but
may be applied to iron (Fe) if we assume the ratio of primary elements to Fe is constant).  
The background-corrected GC color distributions for the entire sample of dE GCC are well-fit
by the Simple model,  as are the GCC distributions for dEs brighter than $M_B = -15$, 
with effective yields $p \sim 10^{-1.5} \Zsun$ ($\bar{\chi}^2 = 0.58-1.38$ ) 
The Simple model does poorly for fainter dEs ($M_B > -15$), although the
numbers are small.  

The comparison of the implied metallicity distributions of the dE globular cluster systems to
Simple closed-box chemical evolution models suggests that a significant amount of metals 
is lost or unavailable for star-formation during the period of globular cluster formation.
In closed systems, the effective yield is expected to reach the true chemical yield of 
supernovae. The low effective yield may be the result of a homogeneous wind, where gas and 
metals are lost proportionally to the star formation rate (Hartwick 1976).
Fainter, less massive dEs would be more likely to experience a terminal wind 
in which all of the galaxy's gas is expelled at a particular time
(for example, via supernovae ``blow-out'';  Arimoto \& Yoshii 1987), 
and results in a sharp cutoff at higher cluster metallicities. 
For such cases where the gas is removed on very short timescales, the
steep metallicity-luminosity scaling relation observed for the 
stellar halos of Local Group dSphs is expected ($Z \propto L^{0.4}$; Dekel \& Silk 1986).   
We are unable to conclusively determine whether the field star colors of our dEs
are more consistent with a $Z \propto L^{0.4}$ and $Z \propto L^{0.2}$ correlation
because of the large scatter in the dE color-magnitude relation.

If the implied correlation between globular cluster metallicity and dE host galaxy 
luminosity is confirmed by more reliable metallicity estimates, and if the metallicity-
luminosity relation of the dE field stars is like that of the Local Group dEs/dSphs, 
then the gas heating mechanism or timescales regulating the chemical 
evolution of dE globular clusters must be markedly different from that regulating the chemical 
evolution of dE field stars.  We note that for slower adiabatic winds 
where the gas is removed on timescales comparable to the crossing time of the system,  
a metallicity - host galaxy luminosity scaling relation similar to that observed for the
globular cluster systems is derived ($Z \propto L^{0.25}$; Dekel \& Silk 1986).
The GC formation model of Kravtsov \& Gnedin (2003), which predicts
too strong of a GC metallicity-host galaxy luminosity relationship, produces GCs during
and after re-ionization in relatively massive halos ($M_{halo} > 10^9 \Msun$).
Smaller halos, however, may be unable to efficiently cool and form the dense molecular
clouds needed for GC production after re-ionization and/or the on-set of supernovae-driven winds. 
GC formation in dwarfs prior to re-ionization would not be subject to the heating effects 
of a high UV-background, and therefore dwarf halos may initially lose metals more 
gradually than at later times.

It is impossible to determine from the $V-I$ colors alone if dE globular 
clusters were formed before or after the epoch of re-ionization.  
Globular clusters formed before re-ionization at redshifts $\geq 7$ would be
$\sim 13$ Gyr old, while clusters formed after the inter-galactic UV background had cooled at 
redshift $\sim 1$ would be $\sim$ 8 Gyr old.
One possible way to distinguish between these two formation epochs is to examine the chemical 
abundance patterns in dE globular clusters. Globular clusters do not have primordial abundances, 
therefore at least some stars must have formed and enriched the gas in dEs before the globular 
clusters were produced. Chemical enrichment of the stellar halo 
would happen quite quickly, as massive stars have lifetimes less than $10^7$ years.   
Assuming stars in the early universe evolve with the same timescales and nuclear 
processes as stars formed at later times (Qian \& Wassberburg 2002), stars in dEs formed 
before the epoch of re-ionization would show enrichment patterns
enhanced in $\alpha$ elements (relative to the Sun) 
created by massive, short-lived stars, as there would not be 
enough time prior to re-ionization for lower mass, primarily iron-producing  
stars to enrich the interstellar medium.  The $\alpha$-element abundance ratios of dE globular clusters 
are unknown, but spectroscopic observations of a small number of dEs suggest that 
their field stars are {\it not} alpha-element 
enhanced (Thomas et al. 2003; Geha et al. 2003), and therefore the majority of dE stars did not form 
prior to re-ionization.

Another possibility is that the dE field stars and globular clusters did not form in separate events,
but rather a combination of strong metallicity gradients in the globular cluster populations and the preferential 
destruction of clusters with higher metallicities has produced the metallicity offset observed today.
This might arise naturally if inner, more metal-rich clusters were destroyed via disk-shocking or
dynamical decay into the dE centers, while outer, more metal-poor clusters migrated inwards. 
However, destruction of more metal-rich clusters via dynamical friction would produce dE nuclei with
colors more like the dE field stars, which is not observed.  
Also, there is little evidence for strong metallicity gradients between
the inner and outer regions of dE.  Most dE show either no color gradients or bluer centers 
(Stiavelli et al. 2001; Jerjen, Binggeli, \& Freeman 2000).  Spectroscopic estimates of
any abundance gradients in the dE field star populations as well as the $\alpha$-element ratios of
dE GCs would help resolve this issue.

\subsection{dE Field Star Formation}
The $V-I$ colors of the dEs are $0.1-0.2$ magnitudes redder than 
their globular clusters, and therefore the majority of
the dE field stars probably did not form in the same event as their globular clusters.
The majority of  dE stars cannot be the remnants of a disrupted GC population (Fall \& Rees 1977),
nor were the Virgo and Fornax dE stellar populations formed in a single
starburst (e.g. Dekel \& Silk 1986).
Given the ages and metallicities derived for nearby dSphs/dEs and their GCs, it is most
likely that the dE stars are more metal-rich, and possibly several Gyr younger 
than their GCS.   An offset in metallicity suggests that the bulk of dE stars were formed after a
period of ISM chemical enrichment, perhaps by the massive stars in the GCs.
The initial epoch of globular cluster formation must have input a significant amount of
energy into the dwarf's remaining gas in the form of UV radiation and supernovae-driven
winds, as would background UV radiation during the epoch of re-ionization (e.g. Babul \& Rees 1992,
Efstathiou 1992).  While the feedback from young clusters and a strong UV-background 
could have halted star-formation temporarily, it must not have been enough to 
completely ``blow-away'' or ``photo-evaporate'' the remaining gas, and therefore
later generations of stars could have formed in the halo once the gas had cooled sufficiently.
The observed solar and sub-solar $\alpha$-element ratios in the dE stellar halos (Geha et al. 2003; 
Thomas et al. 2003) require field star-formation during or after a period of extended 
chemical enrichment, implying either a $>$ 1 Gyr age difference between 
the dE field stars and their GCs or a $>$ 1 Gyr age
spread in the dE field stars.  As we discussed above, the difference between the 
metallicity-host galaxy luminosity relations observed for the dE GCs and field stars suggests that 
different mechanisms regulated the chemical enrichment during each star-formation episode. 
Later star-formation events must have been much less efficient at 
forming massive, tightly bound GCs which could survive to present-day. (Only 
a few intermediate-age GCs have been found in Local Group dSph; van den Bergh 2000.)
The different star-formation regulating mechanisms may have also lowered the 
star-formation efficiency below that required to create gravitationally-bound clusters 
(Kroupa et al. 2001).

We find evidence for a marginally significant color-luminosity trend in the dE,N 
that is not detected in dEs without nuclei.  The non-nucleated dEs also have slightly
bluer $V-I$ colors. This is suggestive of younger ages for the 
non-nucleated dE, although spectroscopically
derived ages for a larger sample will be needed to confirm this.
We find no clear environmental dependence on the dEs and dE,Ns colors (Figure 7).
Other authors have found marginal trends for redder colors with larger projected cluster
distances with ground-based data (Rakos et al. 2001; Secker 1996).  
A lack of correlation between dE color and cluster
position would rule out the popular ``infall'' model for the origin of cluster dEs
(Moore, Lake, \& Katz 1998; Conselice et al. 2003), unless a age-metallicity conspiracy existed in which the
dEs in the outskirts of clusters were both younger and more metal-rich than inner dEs.  
                    
\subsection{dE Nuclei Formation}
The nuclei of dEs are significantly bluer than their associated stellar halos, unlike the 
nuclei and bulges of spirals and giant ellipticals.
The nuclei have colors similar to the globular cluster systems and are as compact
as the globular cluster candidates. They could possess some of the oldest stars in dEs, 
formed first in the densest regions of the dwarf
halos or from the infall of massive globular clusters into the dE centers via dynamical friction
(Lotz et al. 2001, Oh \& Lin 2001, Hernandez \& Gilmore 1998).  Both nuclear formation 
scenarios are consistent with the correlations of 
nuclear color and luminosity with host galaxy luminosity.   Brighter dEs have deeper central
potential wells, and thus would be better able to accrete and retain gas to form brighter and more metal-rich nuclei. 
Brighter dEs also have redder and larger globular cluster populations from which their
nuclei could coalesce.  

Two dEs possess very blue nuclei which require ages $< 2$~Gyr, as does NGC 205, a dE satellite
of M31 (Jones et al. 1996). If many nuclei are relatively young compared to typical globular clusters, 
they could trace a third star-formation episode, after both the globular cluster 
and dE halo formation events.   The trend of bluer nuclei in fainter dE hosts might then imply 
that the nuclei in fainter dEs are typically younger than the nuclei in brighter dEs.  
However, the two very blue nuclei are in dEs brighter than $M_B = -15.8$.
Spectroscopic line indices are needed to place stronger constraints on the
ages, metallicities, and formation scenarios for the dE nuclei.

We find no significant trends for nuclear color or luminosity with local environment
(Figure 8), nor do we find any significant differences between the Virgo and
Fornax dE nuclei.  This is contrary to the IGM pressure-induced star-formation model of Babul \& Rees (1992), 
which predicts that nuclei formed in the densest regions of the cluster undergo the most intense
and rapid starbursts and should be significantly older, more metal-rich and redder than dE nuclei 
formed in the outskirts of the cluster. This also contradicts the infall model for dE formation, 
in which nucleated dEs are the stripped remnants of disk galaxies that experience a central starburst upon
accretion (Moore, Lake, \& Katz 1998). Conversely, the nuclei are typically bluer and 
fainter than the ultra-compact galaxies 
observed in the Fornax Cluster with $-12 < M_V <  -14$ (Karrick et al. 2003). 
While it seems unlikely that the ultra-compact galaxies have been drawn from the 
present-day dE,N population, perhaps the ultra-compacts are consistent with a tidally-disrupted
population of spirals. 

\section{SUMMARY}
We have examined the HST WFPC2 $V-I$ colors for the globular cluster systems, nuclei, and
stellar envelopes of 69 dwarf elliptical galaxies in the Leo Group and Virgo and Fornax Clusters.  
We find the following:

(1) The mean and dispersion in GC $V-I$ color is a function of host dE luminosity.
  Assuming the majority of dE GCs are as old as Local Group dSph GCs, the implied metallicity-luminosity 
  relation is $\langle Z_{GC}\rangle \propto L^{(0.15-0.22)}$,  which is 
  significantly shallower than that observed for the 
  field stars of Local Group dSph ($\langle Z_{FS}\rangle \propto L^{0.4}$).

 (2) The $\langle V-I \rangle_{GC}$ and host galaxy luminosities of E/S0s are consistent 
  with the slope computed for our dE sample plus literature values for 8 local dE/dSph GCS. 
  However, the E/S0 GCS color distributions often
  have a bi-modality not observed in dEs and show no correlation between $\sigma (V-I)$ and galaxy
  luminosity.   The dE GCs are as blue as the metal-poor GC sub-populations observed in the Milky Way
  and local E/S0s. The blue GC peak color - host galaxy luminosity trend for E/S0 reported by 
  Larsen et al. (2001) is consistent with the slope fit to the peak GC colors for our sample of dEs, 
  but is shallower than the slope fit to our sample and  literature dE/dSph GC values.

(3) With the exception of two very blue nuclei, the dE nuclei have colors similar to but slightly
   redder than the GCs, and show the same correlations with host galaxy luminosity.  
  Their colors are also correlated with host galaxy color, as well as with nuclear luminosity.   

(4) We measure the integrated $V-I$ colors for 45 dEs brighter than $M_B = -12.7$.
 The dE field stars are $0.1-0.2$ magnitudes redder than their GCs and nuclei, implying a more metal-rich
 stellar population.  The colors of the nucleated dEs are weakly correlated with their luminosity, while the
 non-nucleated dEs show no significant correlation. 

 (5)  We find no correlation between dE field star color or nuclear color and the dE's projected
 distance from the brightest cluster galaxy, nor do we find any significant difference between
  the colors of the Virgo and Fornax dE stellar populations.

  Therefore, we conclude that: 

   (1) Virgo and Fornax Cluster dEs have probably undergone multiple star-formation episodes. 
       Given the spectroscopic
       metallicities of Local Group dEs/dSphs, it is most likely that the redder dE field 
       star populations are 
       significantly more metal-rich and possibly younger than their associated GCs.  
       At least two dE nuclei trace a third star-formation event less than 1 Gyr ago, after both the
       GC and field star formation epochs.

   (2) dE GCs must have formed within the gravitational potential wells of proto-dE halos, 
       and were not produced in ``pre-galactic'' gas clouds.

   (3) Assuming dE field star populations follow the same metallicity-luminosity relation as 
       Local Group dEs/dSphs, 
       the mechanisms regulating the chemical evolution of dE GCs are significantly different from
       that of the dE field stars.  We suggest that the majority of dE GCs formed at 
       early times, prior to re-ionization and/or strong feedback from supernovae, and therefore lost 
       metals more gradually than stellar populations formed at later times.  
       Subsequent star-formation events (after re-ionization and/or
       the onset of supernovae-driven winds) were subject to more rapid metal-loss
       and are unable to proceed as efficiently as needed to produce significant numbers of GCs.

   (4) Cluster dE evolution models which predict strong color gradients with projected cluster radius for
       stellar envelopes and nuclei are ruled out.  This includes both the infall
       model (Moore, Lake, \& Katz 1998) and IGM pressure-induced star-formation scenario
       (Babul \& Rees 1992).  

This work is based on observations made by the Hubble Space Telescope GO Programs 6352, 7377, and 8500. 
We would like to thank R. Wyse and J. Strader for useful discussions, and 
our anonymous referee for their helpful comments.
J.M.L. acknowledges support from NASA, through LTSA grant number NAG5-11513 
and NASA grant number 9515 from the 
Space Telescope Science Institute, which is operated by AURA, under NASA contract NAS 5-26555.
B.W.M. is supported by the Gemini Observatory, which is operated by the
Association of Universities for Research in Astronomy, Inc., on behalf
of the international Gemini partnership of Argentina, Australia, Brazil,
Canada, Chile, the United Kingdom, and the United States of America.

\begin{deluxetable}{lllllrrcc}
\tablecolumns{9}
\tablewidth{0pc}
\tabletypesize{\scriptsize}
\tablecaption{HST dE Snapshot Survey I, II, \& III \label{de_tab}}
\tablehead{ \colhead{Galaxy\tablenotemark{1}} & \colhead{IDs\tablenotemark{2}}
 & \colhead{$B$\tablenotemark{1}} & \colhead{$M_B$\tablenotemark{3}} 
& \colhead{Type\tablenotemark{1}}  &\colhead{RA(J2000)} & \colhead{ DEC(J2000)}
& \colhead{$A_V$\tablenotemark{4}} & \colhead{$E_{VI}$\tablenotemark{4}} }
\startdata
LGC47    &  UGC05944         &     15.2 &  $-15.1$ &  dE        & 10:50:18  & 13:16:19   &0.093 &0.039\\
LGC50    &  \nodata          &     17.0 &  $-13.3$ &  dEN       & 10:51:01  & 13:20:02   &0.093 &0.039\\
\cutinhead{}
FCC25    &  NG31	     &     17.7 &  $-13.7$ &  dE0N	&  3:23:33 & $-$36:58:57 &   0.069 &    0.029\\
FCC27    &  \nodata	     &     19.3 &  $-12.1$ &  dE2	&  3:23:56 & $-$34:12:29 &   0.035 &    0.014\\
FCC46    &  \nodata	     &     15.6 &  $-15.8$ &  dE4      &   3:26:25 & $-$37.53:20 &   0.061 &    0.025\\
FCC48    &  NG61             &     17.1 &  $-14.3$ &  dE3       &  3:26:42 & $-$34:28:04  &0.036 &0.015 \\
FCC59    &  \nodata	     &     19.4 &  $-12.0$ &  dE0N	&  3:27:47 & $-$33:27:10 &   0.043 &    0.018\\
FCC64    &  \nodata          &     17.5 &  $-13.9$ &  dE5       &  3:38:00  & $-$38:28:47  &0.061 &0.025\\
FCC110   &  NG26, ESO358-G14  &     16.8 &  $-14.6$ &  dE4       &  3:32:57  & $-$35:16:44  &0.050 &0.021\\
FCC136   &  NG24             &     14.8 &  $-16.6$ &  dE2N      &  3:34:29  & $-$35:28:11  &0.053 &0.022\\
FCC144   &  \nodata	     &     19.2 &  $-12.2$ &  dE0	&  3:35:00 &	  $-$35:41:39 &   0.047 &    0.020\\
FCC146   &  \nodata	     &     19.5 &  $-11.9$ &  dE4N	&  3:35:31 &	  $-$35:41:37 &   0.049 &    0.021\\
FCC150   &  \nodata          &     15.7 &  $-15.7$ &  dE4       &  3:35:24  & $-$36:39:10  &0.046 &0.019\\
FCC174   &  \nodata          &     16.7 &  $-14.7$ &  dE1N      &  3:36:45  & $-$33:00:10  &0.026 &0.011\\
FCC189   &  \nodata	     &     18.8 &  $-12.6$ &  dE4N	&  3:37:08 &	  $-$34:17:08 &   0.046 &    0.019\\
FCC212   &  NG119	     &     17.6 &  $-13.8$ &  dE1?     &   3:38:21 &     $-$36:36:12 &   0.036 &    0.015\\
FCC218   &  NG72  	     &     18.5 &  $-12.9$ &  dE4      &   3:38:45 &     $-$35:41:31 &   0.041 &    0.017\\
FCC238   &  NG109	     &     18.7 &  $-12.7$ &  dE5N     &   3:40:17 & 	$-$36:28:54 &   0.031 &    0.013\\
FCC242   &  NG51	     &     17.8 &  $-13.6$ &  dE5	&  3:40:20 &	  $-$37:22:21 &   0.047 &    0.020\\
FCC246   &  \nodata	     &     19.1 &  $-12.3$ &  dE2	&  3:40:37 &	  $-$36:53:43 &   0.038 &    0.016\\
FCC254   &  NG22             &     17.6 &  $-13.8$ &  dE0N      &  3:41:00  & $-$35:16:26  &0.034 &0.014\\
FCC304   &  \nodata	     &     18.8 &  $-12.6$ &  dE1	&  3:45:30 &	 $-$34:30:39 &   0.023 &    0.010\\
FCC316   &  NG12             &     16.7 &  $-14.7$ &  dE1N      &  3:36:45  & $-$33:00:10  &0.036 &0.015\\
FCC324   &  ESO358-G66       &     15.3 &  $-16.1$ &  dS01(8)   &  3:47:52  & $-$36:32:41  &0.032 &0.013\\
\cutinhead{}
VCC9     &  IC3019, UGC07136  &     13.9 &  $-17.0$ &  dE1N     &  12:09:22  & 13:59:34  &0.128 &0.053\\
VCC118   &  R92 	     &     15.6 &  $-15.3$ &  dE3      &  12:14:36  & 9:41:20   &0.053 &0.022\\
VCC128   &  UGCA275, R903     &     15.6 &  $-15.3$ &  dE0      &  12:14:59 & 	  9:33:54 &   0.051 &    0.022\\
VCC158   &  UGC07269, R1513   &     15.8 &  $-15.1$ &  dE3N     &  12:15:40 & 	15:00:18 &   0.131 &    0.055\\
VCC240   &  R1419	     &     18.2 &  $-12.7$ &  dE2N     &  12:17:31  & 14:21:21  &0.108 &0.045\\
VCC452   &  R1223	     &     15.8 &  $-15.1$ &  dE4N     &  12:21:04  & 11:45:18  &0.093 &0.039\\
VCC503   &  R817	     &     16.8 &  $-14.1$ &  dE3N     &  12:21:50  &  8:32:31  &0.062 &0.026\\
VCC529   &  R1015	     &     18.2 &  $-12.7$ &  dE4N     &  12:22:08 & 	 9:53:41 &   0.079 &    0.033\\
VCC543   &  UGC07436	     &     14.8 &  $-16.1$ &  dE5      &  12:22:19 & 	14:45:38 &   0.105 &    0.045\\
VCC546   &  R1017	     &     15.7 &  $-15.2$ &  dE6      &  12:22:21   & 	10:36:06 &   0.121 &    0.050\\
VCC646   &  \nodata	     &     18.8 & $-12.1$ &  dE3       &  12:23:31 & 	 17:47:42 &   0.099 &    0.041\\
VCC747   &  R924	     &     16.2 &  $-14.7$ &  dE0N     &  12:24:47 & 	 8:59:21 &   0.072 &    0.030\\
VCC871   &  R1240	     &     15.4 &  $-15.5$ &  dE4N     &  12:26:05 & 	12:33:33 &   0.109 &    0.045\\
VCC896   &  R1352	     &     17.8 &  $-13.1$ &  dE3N     &  12:26:22 & 	12:47:03 &   0.096 &    0.040\\
VCC917   &  IC3344	     &     14.9 &  $-16.0$ &  dE6      &  12:26:32  & 13:34:44  &0.107 &0.045\\
VCC940   &  IC3349, R1243     &     14.8 &  $-16.1$ &  dE1N     &  12:26:47 & 	12:27:12 &   0.101 &    0.042\\
VCC949   &  R1027	     &     15.1 &  $-15.8$ &  dE4N     &  12:26:54 & 	10:39:56 &   0.102 &    0.042\\
VCC965   &  IC3363, R1245     &     15.4 &  $-15.5$ &  dE7N     &  12:27:03 & 	12:33:37 &   0.100 &    0.042\\
VCC992   &  R827	     &     15.8 &  $-15.1$ &  dE0N     &  12:27:18 & 	 8:12:45 &   0.075 &    0.031\\
VCC996   &  R1360	     &     18.4 &  $-12.5$ &  dE5      &  12:27:21 & 	13:06:41 &   0.100 &    0.041\\
VCC1073  &  IC794, UGC07585   &     14.2 &  $-16.7$ &  dE3N     &  12:28:08  & 12:05:34  &0.091 &0.038\\
VCC1077  &  \nodata	     &     19.2 &  $-11.7$ &  dE0N     &  12:28:09 & 	12:48:23 &   0.078 &    0.032\\
VCC1105  & R1440	     &     16.2 &  $-14.7$ &  dE0N     &  12:28:27 & 	14:09:15 &   0.146 &    0.061\\
VCC1107  &  NGC4472 DW01, R731 &     15.1 &  $-15.8$ &  dE4N     &  12:28:30 & 	 7:19:28 &   0.069 &    0.029\\
VCC1252  &  \nodata	     &     18.8 &  $-12.1$ &  dE0N     &  12:30:01 & 	 9:28:24 &   0.066 &    0.028\\
VCC1254  &  NGC4472 DW08, R834 &     15.0 &  $-15.9$ &  dE0N    &  12:30:05  & 8:04:22 &0.074 &0.031\\
VCC1272  &  \nodata	     &     18.5 &  $-12.4$ &  dE1N     &  12:30:15 & 	13:18:28 &   0.088 &    0.036\\
VCC1308  &  IC3437	     &     15.1 &  $-15.8$ &  dE6N     &  12:30:46 & 	11:20:33 &   0.116 &    0.048\\
VCC1311  &  NGC4472 DW03, R737 &     15.6 &  $-15.3$ &  dE1N     &  12:30:47 & 	  7:36:18 &   0.063 &    0.026\\
VCC1363  &  \nodata	     &     19.0 &  $-11.9$ &  dE3N     &  12:31:27  & 	10:56:07 &   0.107 &    0.044\\
VCC1386  &  IC3457, UGC3457, R1264  &  14.4 &  $-16.5$ &  dE3N   &  12:31:51 & 	12:39:42 &   0.084 &    0.035\\
VCC1497  &  UGC07707	     &     15.7 &  $-15.2$ &  dE4N     &  12:33:18 & 	17:27:35 &   0.086 &    0.036\\
VCC1514  &  \nodata	     &     15.1 &  $-15.8$ &  dE7N     &  12:33:37 & 	 7:52:14 &   0.064 &    0.026\\
VCC1530  &  \nodata	     &     18.3 &  $-12.6$ &  dE2N     &  12:33:55 & 	 5:43:07 &   0.075 &    0.031\\
VCC1577  &  IC3519, R1555     &     15.8 &  $-15.1$ &  dE4      &  12:34:38  & 15:36:10  &0.093 &0.038\\
VCC1651  &  R640	     &     17.0 &  $-13.9$ &  dE5      &  12:36:07  &  6:03:10  &0.063 &0.026\\
VCC1714  &  R1461	     &     18.5 &  $-12.4$ &  dE4N     &  12:37:25 & 	 14:18:46 &   0.097 &    0.040\\
VCC1729  &  \nodata	     &     17.8 &  $-13.1$ &  dE5?     &  12:37:46 & 	10:59:07 &   0.077 &    0.032\\
VCC1762  &  R1055	     &     16.2 &  $-14.7$ &  dE6      &  12:38:32  & 10:22:37  &0.072 &0.030\\
VCC1781  &  R842	     &     18.7 &  $-12.2$ &  dE4      &  12:39:11 & 	 8:04:17 &   0.087 &    0.036\\
VCC1876  &  IC3658, R1465     &     14.9 &  $-16.0$ &  dE5N     &  12:41:20  & 14:42:03  &0.103 &0.043\\
VCC1877  &  R844	     &     18.6 &  $-12.3$ &  dE2      &  12:41:23 & 	 8:22:00 &   0.093 &    0.038\\
VCC1948  &  R1062	     &     15.1 &  $-15.8$ &  dE3      &  12:42:58 & 	10:40:54 &   0.080 &    0.033\\
VCC2008  &  IC3720, DDO145    &     15.1 &  $-15.8$ &  dE5      &  12:44:46 & 	 12:03:57 &   0.090 &    0.037\\
VCC2029  &  R964	     &     18.2 &  $-12.7$ &  dE3      &  12:45:40  &  9:24:25  &0.082 &0.034
\enddata
\tablenotetext{1}{FCC = Fornax Cluster Catalog, Ferguson 1989; VCC = Virgo Cluster Catalog, Binggeli, Sandage \& Tammann 1985; LGC = 
Leo Group Catalog, Ferguson \& Sandage 1990; apparent B magnitude and morphological types taken from these catalogs.}
\tablenotetext{2}{NG = New Galaxy, IC = Index Catalog, UGC = Uppsala General Catalog of Galaxies, NGC= New General Catalog, 
R = Reaves 1983, DDO = David Dunlop Observatory Catalogue of Low Surface Brightness Galaxies}
\tablenotetext{3}{Absolute B magnitudes calculated assuming $(M-m)$ = 30.0 for Leo Group, 30.92 for Virgo Cluster, and 31.39 for
the Fornax Cluster (Freedman et al. 2001).  The values have not been corrected for foreground extinction.}
\tablenotetext{4}{ The Schlegel et al. 1998 values for foreground Galactic extinction, taken from the NASA/IPAC Extragalactic Database (NED),
which is operated by the Jet Propulsion Laboratory, California Institute of Technology, under contract with the National 
Aeronautics and Space Administration.}
\end{deluxetable}
\clearpage

\clearpage
\begin{deluxetable}{lllrrrrrrr}
\tablecolumns{10}
\tablewidth{0pc}
\tabletypesize{\scriptsize}
\tablecaption{Nucleated Dwarf Elliptical Galaxies \label{nde_tab}}
\tablehead{ \colhead{Galaxy} & \colhead{$M_B$} & \colhead{$V$\tablenotemark{1}} 
& \colhead{$V-I$} & \colhead{$\epsilon$} &\colhead{$r_0$ (\arcsec)} 
&\colhead{$r_0$ (kpc)} &\colhead{$M_V$(nucleus)} &\colhead{$V-I$ (nucleus)} 
&\colhead{$N_{GC}$ \tablenotemark{2}}}
\startdata
VCC1073 &  $-16.7$  &  13.89 $\pm$  0.11 & 1.15   $\pm$  0.03 &    0.30 &     8.6 $\pm$   0.5 &   636.6 &  $-11.24$  $\pm$  0.01 &  1.07  $\pm$   0.02 &   16.7  $\pm$  5.8   \\
FCC136  &  $-16.6$  &  14.32 $\pm$  0.18 & 1.12   $\pm$  0.03 &    0.20 &     8.4 $\pm$   1.3 &   641.9 &  $-11.40$  $\pm$  0.01 &  1.05  $\pm$   0.02 &   18.0  $\pm$  5.3  \\
VCC1386 &  $-16.5$  &   \nodata          & 1.06   $\pm$  0.03 &    0.30 &    12.7 $\pm$   0.1 &   945.3 &  $-10.73$  $\pm$  0.04 &  0.88  $\pm$   0.04 &   22.7  $\pm$  6.1  \\
VCC940  &  $-16.1$  &   \nodata          & 1.08   $\pm$  0.03 &    0.16 &    10.8 $\pm$   0.2 &   799.9 &  $-11.16$  $\pm$  0.04 &  0.96  $\pm$   0.04 &   52.4  $\pm$  8.1  \\
FCC324 &   $-16.1$  &  14.95 $\pm$  0.10 & 1.10   $\pm$  0.03 &    0.63 &    13.9 $\pm$   0.3 &  1283.9 &   $-8.80$  $\pm$  0.05 &  0.91  $\pm$  0.05 &     9.0  $\pm$  4.6  \\
VCC1876 &  $-16.0$  &  14.39 $\pm$  0.18 & 1.04   $\pm$  0.03 &    0.49 &    10.5 $\pm$   0.3 &   780.6 &  $-10.40$  $\pm$  0.03 &  0.97  $\pm$  0.02 &    22.1  $\pm$  6.3  \\
VCC1254 &  $-15.9$  &  14.82 $\pm$  0.13 & 1.16   $\pm$  0.03 &    0.03 &     7.7 $\pm$   0.2 &   568.4 &  $-12.25$  $\pm$  0.02 &  1.01  $\pm$  0.02 &    20.5  $\pm$  8.0  \\
VCC949  &  $-15.8$  &   \nodata          & 1.11   $\pm$  0.03 &    0.30 &    10.7 $\pm$   0.3 &   793.2 &   $-9.80$  $\pm$  0.04 &  0.93  $\pm$  0.04 &    21.6  $\pm$  5.5  \\
VCC1308 &  $-15.8$  &   \nodata          & 1.05   $\pm$  0.03 &    0.30 &     6.0 $\pm$   0.4 &   443.7 &  $-11.41$  $\pm$  0.04 &  1.05  $\pm$  0.04 &    19.7  $\pm$  5.9  \\
VCC1514 &  $-15.8$  &   \nodata          & 1.00   $\pm$  0.03 &    0.67 &    12.6 $\pm$   0.3 &   932.0 &   $-8.89$  $\pm$  0.04 &  0.92  $\pm$  0.05 &     7.0  $\pm$  4.4  \\
VCC1107 &  $-15.8$  &   \nodata          & 1.12   $\pm$  0.05 &    0.33 &    11.8 $\pm$   0.3 &   871.8 &  $-10.32$  $\pm$  0.04 &  0.96  $\pm$   0.03 &    1.8  $\pm$  2.5  \\
FCC46   &  $-15.8$  &   \nodata          & 0.88   $\pm$  0.03 &    0.38 &     5.3 $\pm$   0.7 &   489.0 &  $-12.16$  $\pm$  0.03 &  0.58  $\pm$   0.04 &    8.0  $\pm$  4.2  \\
FCC150 &   $-15.7$  &  15.13 $\pm$  0.10 & 1.05   $\pm$  0.03 &    0.15 &     5.5 $\pm$   0.3 &   502.9 &  $-11.01$  $\pm$  0.02 &  1.03  $\pm$   0.02 &    8.0  $\pm$  4.2  \\
VCC871  &  $-15.5$  &   \nodata          & 1.05   $\pm$  0.03 &    0.33 &    10.8 $\pm$   0.2 &   785.8 &   $-9.10$  $\pm$  0.03 &  0.88  $\pm$   0.05 &   15.6  $\pm$  4.9  \\
VCC965  &  $-15.5$  &   \nodata          & 1.02   $\pm$  0.03 &    0.49 &    13.4 $\pm$   0.3 &   990.6 &  $-10.97$  $\pm$  0.04 &  0.92  $\pm$   0.03 &   13.2  $\pm$  4.6  \\
VCC1311 &  $-15.3$  &   \nodata          & 1.03   $\pm$  0.03 &    0.08 &     9.6 $\pm$   0.3 &   713.8 &  $-10.55$  $\pm$  0.03 &  0.90  $\pm$   0.04 &   15.7  $\pm$  4.7  \\
VCC1497 &  $-15.2$  &   \nodata          & 1.04   $\pm$  0.03 &    0.47 &    12.7 $\pm$   0.5 &   945.3 &   $-9.13$  $\pm$  0.05 &  0.89  $\pm$   0.05 &    5.0  $\pm$  4.6  \\
VCC158  &  $-15.1$  &   \nodata          & 1.14   $\pm$  0.03 &    0.33 &    12.5 $\pm$   0.3 &   929.0 &  $-10.08$  $\pm$  0.05 &  0.95  $\pm$   0.05 &    3.0  $\pm$  4.1  \\
VCC452 &   $-15.1$  &  15.36 $\pm$  0.11 & 1.06   $\pm$  0.03 &    0.22 &     7.8 $\pm$   0.4 &   579.5 &   $-8.50$  $\pm$  0.04 &  0.91  $\pm$   0.06 &   15.5  $\pm$  5.3  \\
VCC992  &  $-15.1$  &   \nodata          & 1.04   $\pm$  0.06 &    0.08 &     7.8 $\pm$   0.3 &   580.2 &   $-9.20$  $\pm$  0.04 &  0.89  $\pm$   0.04 &   12.5  $\pm$  5.0  \\
VCC1577 &  $-15.1$  &  15.17 $\pm$  0.31 & 1.04   $\pm$  0.03 &    0.24 &     7.1 $\pm$   0.4 &   523.9 &   $-8.64$  $\pm$  0.05 &  0.90  $\pm$   0.04 &    8.9  $\pm$  4.6  \\
FCC174 &   $-14.7$  &  16.16 $\pm$   0.1 & 0.97   $\pm$  0.08 &    0.17 &     4.9 $\pm$   0.5 &   454.1 &   $-9.76$  $\pm$  0.02 &  0.92  $\pm$   0.03 &    4.1  $\pm$  3.4  \\
FCC316 &   $-14.7$  &  16.17 $\pm$   0.1 & 1.06   $\pm$  0.03 &    0.30 &     8.2 $\pm$   0.3 &   750.6 &   $-9.49$  $\pm$  0.02 &  1.07  $\pm$   0.03 &    8.0  $\pm$  3.7  \\
VCC1105 &  $-14.7$  &   \nodata          & 0.93   $\pm$  0.17 &    0.14 &    10.3 $\pm$   0.5 &   761.3 &   $-9.82$  $\pm$  0.04 &  0.87  $\pm$   0.04 &   14.0  $\pm$  4.9  \\
VCC747  &  $-14.7$  &   \nodata          &  \nodata           &    0.20 &    10.1 $\pm$   0.7 &   747.2 &  $-10.02$  $\pm$  0.03 &  0.88  $\pm$   0.04 &   17.0  $\pm$  5.6  \\
VCC503 &   $-14.1$  &  16.83 $\pm$  0.10 & 1.02   $\pm$  0.15 &    0.06 &     6.3 $\pm$   0.4  &  468.2 &   $-9.06$  $\pm$  0.03 &  0.91  $\pm$   0.03 &    2.5  $\pm$  3.7  \\
FCC254 &   $-13.8$  &  16.71 $\pm$  0.21 & 1.07   $\pm$  0.10 &    0.16 &    10.5 $\pm$   0.5 &   967.1 &  $-10.58$  $\pm$  0.01 &  0.91  $\pm$   0.02 &    4.5  $\pm$  4.2  \\
FCC25   &  $-13.7$  &   \nodata          &  \nodata           &    0.25 &     6.1 $\pm$   0.3 &   564.6 &   $-9.88$  $\pm$  0.03 &  0.90  $\pm$   0.04 &    2.5  $\pm$  3.7  \\
FCC242 &   $-13.6$  &   \nodata          &  \nodata           &    0.43 &     9.3 $\pm$   0.6 &   852.8 &   $-8.22$  $\pm$  0.05 &  0.94  $\pm$   0.07 &    0.0  $\pm$  2.7  \\
LGC50 &    $-13.3$  &  16.04 $\pm$  0.21 & 1.03   $\pm$  0.03 &    0.36 &    10.2 $\pm$   0.4 &   494.2 &   $-8.50$  $\pm$  0.02 &  0.80  $\pm$   0.03 &    7.2  $\pm$  3.8  \\
VCC896  &  $-13.1$  &   \nodata          & 0.98   $\pm$  0.15 &    0.38 &     8.8 $\pm$   0.6 &   655.9 &   $-8.35$  $\pm$  0.04 &  0.99  $\pm$   0.06 &    4.5  $\pm$  3.9  \\
VCC529 &   $-12.7$  &   \nodata          &  \nodata           &    0.43 &     6.6 $\pm$   0.5 &   489.7 &   $-9.67$  $\pm$  0.03 &  0.88  $\pm$   0.04 &    2.0  $\pm$  2.0  \\
FCC238  &  $-12.7$  &   \nodata          &  \nodata           &    0.34 &     7.0 $\pm$   0.7 &   644.7 &   $-9.32$  $\pm$  0.03 &  0.89  $\pm$   0.05 &    7.2  $\pm$  4.1  \\
VCC240 &   $-12.7$  &  16.34 $\pm$  0.70 &  0.98   $\pm$ 0.10 &    0.35 &     9.5 $\pm$   1.0 &   705.6 &   $-9.46$  $\pm$  0.02 &  0.91  $\pm$   0.03 &    3.5  $\pm$  3.5  \\
FCC189 &   $-12.6$  &   \nodata &  \nodata                    &    0.26 &     2.4 $\pm$   0.3 &   222.0 &   $-9.17$  $\pm$  0.03 &  0.81  $\pm$   0.05 &    6.5  $\pm$  4.2  \\
VCC1530 &  $-12.6$  &   \nodata &  \nodata                    &    0.15 &     5.9 $\pm$   0.3 &   440.7 &   $-9.73$  $\pm$  0.03 &  0.84  $\pm$   0.04 &    0.5  $\pm$  3.4  \\
VCC1714 &  $-12.4$  &   \nodata &  \nodata                    &    0.39 &    12.0 $\pm$   0.8 &   893.4 &  $-11.82$  $\pm$  0.02 &  0.55  $\pm$   0.03 &    2.5  $\pm$  3.4  \\
VCC1272 &  $-12.4$  &   \nodata &  \nodata                    &    0.40 &     4.2 $\pm$   0.4 &   308.7 &   $-7.66$  $\pm$  0.05 &  0.69  $\pm$   0.09 &    0.0  $\pm$  3.8  \\
FCC246 &   $-12.3$  &   \nodata &  \nodata                    &    0.38 &     6.6 $\pm$   0.7 &   609.7 &   $-7.28$  $\pm$  0.09 &  0.84  $\pm$   0.13 &    0.0  $\pm$  2.7  \\
VCC646 &   $-12.1$  &   \nodata &  \nodata                    &    0.23 &     5.6 $\pm$   0.2 &   414.1 &   $-9.55$  $\pm$  0.03 &  0.96  $\pm$   0.04 &    6.5  $\pm$  4.2  \\
VCC1252 &  $-12.1$  &   \nodata &  \nodata                    &    0.05 &     4.7 $\pm$   0.5 &   345.0 &   $-9.02$  $\pm$  0.03 &  0.86  $\pm$   0.04 &    4.5  $\pm$  3.9  \\
FCC59 &    $-12.0$  &   \nodata &  \nodata                    &    0.38 &    11.0 $\pm$   2.2 &  1016.8 &   $-8.72$  $\pm$  0.04 &  0.83  $\pm$   0.06 &    0.0  $\pm$  3.2  \\
VCC1363 &  $-11.9$  &   \nodata &  \nodata                    &    0.35 &     3.7 $\pm$   0.2 &   271.6 &   $-8.83$  $\pm$  0.03 &  0.89  $\pm$   0.05 &    0.5  $\pm$  2.7  \\
FCC146 &   $-11.9$  &   \nodata &  \nodata                    &    0.49 &     3.3 $\pm$   0.4 &   304.9 &   $-8.11$  $\pm$  0.05 &  1.01  $\pm$   0.08 &    0.0  $\pm$  2.6  \\
VCC1077 &  $-11.7$  &   \nodata &  \nodata                    &    0.18 &     3.7 $\pm$   0.3 &	242.6 &     $-8.45$  $\pm$  0.04 &  0.93  $\pm$   0.05 &    1.5  $\pm$  4.5  \\
\enddata
\tablenotetext{1}{From Stiavelli et al. 2001}
\tablenotetext{2}{Corrected for background contamination}
\end{deluxetable}
\clearpage

\begin{deluxetable}{lllllrrr}
\tablecolumns{8}
\tablewidth{0pc}
\tabletypesize{\scriptsize}
\tablecaption{Non-nucleated Dwarf Elliptical Galaxies \label{nonde_tab}}
\tablehead{ \colhead{Galaxy} & \colhead{$M_B$} & \colhead{$V$\tablenotemark{1}} & \colhead{$V-I$} 
& \colhead{$\epsilon$} &\colhead{$r_0$ (\arcsec)} 
&\colhead{$r_0$(kpc)} &\colhead{$N_{GC}$\tablenotemark{2}}}
\startdata
VCC9     &  $-17.0$ &  13.21 $\pm$ 0.10  &  1.00 $\pm$   0.03  & 0.16  & 18.3 $\pm$ 0.6 & 1358.6  &  23.9   $\pm$   6.9    \\
VCC543   &  $-16.1$ &  \nodata           &  1.07 $\pm$   0.03  & 0.46  &  8.5 $\pm$ 0.7 &  630.0  &  13.0   $\pm$   4.4   \\
VCC917   &  $-16.0$ &  14.76 $\pm$ 0.10  &  1.01 $\pm$   0.03  & 0.46  &  6.0 $\pm$ 0.2 &  443.0  &   6.0   $\pm$   5.1   \\
VCC1948  &  $-15.8$ &   \nodata          &  0.99 $\pm$   0.03  & 0.28  &  7.3 $\pm$ 0.1 &  540.9  &   6.5   $\pm$   3.7   \\
VCC2008  &  $-15.8$ &   \nodata          &  1.00 $\pm$   0.03  & 0.71  & 36.8 $\pm$ 0.9 & 2729.1  &   7.5   $\pm$   5.1   \\
VCC118   &  $-15.3$ &  15.55 $\pm$ 0.15  &  0.94 $\pm$   0.03  & 0.22  &  8.0 $\pm$ 0.6 &  592.2  &   3.0   $\pm$   3.6   \\
VCC128   &  $-15.3$ &   \nodata          &  1.10 $\pm$   0.05  & 0.09  & 12.1 $\pm$ 0.6 &  898.6  &  11.1   $\pm$   4.4   \\
VCC546   &  $-15.2$ &   \nodata          &  0.91 $\pm$   0.03  & 0.61  & 10.6 $\pm$ 0.4 &  785.8  &   4.8   $\pm$   3.3   \\
LGC47    &  $-15.1$ &  14.57 $\pm$ 0.10  &  1.02 $\pm$   0.03  & 0.13  & 12.9 $\pm$ 0.6 &  623.2  &   2.9   $\pm$   4.4   \\
VCC1762  &  $-14.7$ &  15.81 $\pm$ 0.20  &  0.97 $\pm$   0.03  & 0.53  &  6.9 $\pm$ 0.2 &  504.6  &   4.0   $\pm$   3.5   \\
FCC110   &  $-14.6$ &  16.49 $\pm$ 0.10  &  1.14 $\pm$   0.03  & 0.40  & 10.8 $\pm$ 0.4 &  994.7  &   0.0   $\pm$   4.2   \\
FCC48    &  $-14.3$ &  16.55 $\pm$ 0.13  &  1.00 $\pm$   0.03  & 0.45  &  6.9 $\pm$ 0.4 &  631.8  &   7.0   $\pm$   4.4   \\
FCC64    &  $-13.9$ &  17.09 $\pm$ 0.10  &  1.12 $\pm$   0.03  & 0.25  &  8.3 $\pm$ 0.7 &  767.2  &   0.0   $\pm$   2.9   \\
VCC1651  &  $-13.9$ &  16.08 $\pm$ 0.29  &  1.11 $\pm$   0.11  & 0.44  & 19.2 $\pm$ 1.1 & 1424.6  &   4.2   $\pm$   5.3   \\
FCC212   &  $-13.8$ &   \nodata          &   \nodata	     & 0.36    & 15.1 $\pm$ 0.8 & 1388.9  &   5.5   $\pm$   5.3   \\
VCC1729  &  $-13.1$ &   \nodata          &  0.98 $\pm$   0.03  & 0.47  &  6.3 $\pm$ 0.4 &  466.0  &   1.0   $\pm$   3.0   \\
FCC218   &  $-12.9$ &   \nodata          &   \nodata           & 0.31  &  5.6 $\pm$ 0.5 &  513.9  &   0.0   $\pm$   2.7   \\
VCC2029  &  $-12.7$ &  16.98 $\pm$ 0.28  &  1.00 $\pm$   0.03  & 0.32  &  5.0 $\pm$ 0.3 &  463.3  &   1.0   $\pm$   3.6   \\
FCC304   &  $-12.6$ &   \nodata &   \nodata &    0.15  &	          7.2 $\pm$ 0.9 &  660.4  &   0.0   $\pm$   3.1   \\
VCC996   &  $-12.5$ &   \nodata &   \nodata &    0.33  &	          7.0 $\pm$ 0.4 &  517.9  &   0.0   $\pm$   3.7   \\
VCC1877  &  $-12.3$ &   \nodata &   \nodata &    0.43  &	          7.4 $\pm$ 0.2 &  546.9  &   3.0   $\pm$   3.3   \\
VCC1781  &  $-12.2$ &   \nodata &   \nodata &    0.32  &	          5.6 $\pm$ 0.4 &  417.7  &   2.5   $\pm$   3.4   \\
FCC144   &  $-12.2$ &   \nodata &   \nodata &    0.19  &	          3.7 $\pm$ 0.4 &  337.1  &   0.5   $\pm$   3.4   \\
FCC27    &  $-12.1$ &   \nodata &   \nodata &    0.26  &	          3.6 $\pm$ 0.5 &  334.3  &   0.5   $\pm$   3.1   \\
\enddata
\tablenotetext{1}{From Stiavelli et al. 2001}
\tablenotetext{2}{Corrected for background contamination.}
\end{deluxetable}

\clearpage
\begin{deluxetable}{rrcrcc}
\tablecolumns{6}
\tablewidth{0pc}
\tabletypesize{\scriptsize}
\tablecaption{ Composite dE Globular Cluster Color Distributions \label{vi_tab}}
\tablehead{ \colhead{ Magnitude Bin} & \colhead{ N$_{galaxy}$} 
& \colhead{$\langle M_B \rangle$}  & \colhead{ N$_{GC}$}  & \colhead{ $\langle V-I \rangle$\tablenotemark{1}}
&\colhead{ $\sigma (V-I)$\tablenotemark{1}} } 
\startdata
$-11.7 > M_B>  -13.0$ & 22 & $-12.4$ & 28 $\pm$ 16 & 0.83 $\pm$ 0.02 & 0.03 $\pm$ 0.01 \\
$-13.0 > M_B >  -14.0$ &  9 & $-13.7$ & 29 $\pm$ 12 & 0.84 $\pm$ 0.05 & 0.04 $\pm$ 0.03 \\
$-14.0 > M_B >  -15.0$ &  8 & $-14.7$ & 54 $\pm$ 10 & 0.91 $\pm$ 0.03 & 0.07 $\pm$ 0.05 \\
$-15.0 > M_B >  -16.0$ & 21 & $-15.5$ & 216 $\pm$ 22 & 0.89 $\pm$ 0.01 & 0.10 $\pm$ 0.02 \\
$-16.0 > M_B >  -17.0$ &  9 & $-16.1$ & 194 $\pm$ 18 & 0.90 $\pm$ 0.01 & 0.13 $\pm$ 0.01 \\
\enddata
\tablenotetext{1}{Median Gaussian fit and standard deviation of fitted parameter for 
10,000 bootstrap realizations of background-corrected GC color distributions.}
\end{deluxetable}

\begin{deluxetable}{lccclllcl}
\setlength{\tabcolsep}{0.05in}
\tablecolumns{9}
\tablewidth{0pc}
\tabletypesize{\scriptsize}
\tablecaption{Literature dE/dSph Globular Cluster System and Field Star Properties \label{lg_gc_tab}}
\tablehead{ \colhead{Galaxy} & \colhead{$M_B$} & \colhead{$M_V$}  
&\colhead{N$_{GC}$} &\colhead{ $\langle$[Fe/H]$\rangle_{GC}$} &\colhead{ $\langle V-I \rangle_{GC}$} 
&\colhead{$\langle $[Fe/H]$\rangle_{FS}$}  &\colhead{$\langle$Age$\rangle_{FS}$} &\colhead{Notes}
}
\startdata
\cutinhead{Local Group dSph/dE}
Fornax      & $ -12.6$  &$-13.1$  & 5    &$-1.8\pm$ 0.2  & 0.83$\pm$0.06    & $-0.9\pm 0.1$ & $\leq$ 4 Gyr  & 1,2,3 \\
Sagittarius &  $-12.8$  &$-13.3$  & 4    &$-1.6\pm$ 0.5  & 0.89$\pm$0.14    & $-0.5\pm$ 0.1 & 6$\pm$2 Gyr   & 1,4,5; includes Terzan7\\
NGC147      &  $-14.8$  &$-15.2$  & 2    &$-2.0\pm$ 0.4  & 0.75$\pm$0.12 & $-0.9\pm$ ?   & $\geq$ 4 Gyr  & 1,6,7\\
NGC185      &  $-14.7 $ &$-15.5$  & 5    &$-1.7\pm$ 0.3  & 0.86$\pm$0.10 & $-1.2\pm$0.2  & $\geq$ 4 Gyr  & 1,6,8,9\\
NGC205      &  $-16.0$  &$-16.6$  & 7    &$-1.4\pm$ 0.1  & 0.95$\pm$0.04 & $-0.8\pm$ 0.2 & $\sim$ 4 Gyr  & 1,6,10 \\
\cutinhead{}
VCC1254     &  $-15.9$  &$-16.1$  & $22 \pm 8$ & $-1.5\pm$ 0.1  & 0.93$\pm$0.06 & $-1.1\pm$ 0.2 & \nodata       & 14\\
VCC1386     &  $-16.5$  &$-16.8$  & $17 \pm 5$ & $-1.5\pm$ 0.2  &0.93$\pm$0.02  & $-0.8\pm$ 0.2  & \nodata       & 14\\
NGC3115DW1  &  $-17.0$  &$-17.7$  & 7/37 &$-1.0\pm$ 0.2  & 0.98$\pm$0.02 & $-0.7\pm$ 0.3 & \nodata       & 11,12,13; $\langle V-I\rangle_{GC}$ from 13 \\
\enddata
\tablecomments{(1) Mateo 1998; (2) Strader et al. 2003; (3) Pont et al. 2004; (4) Montegriffo et al. 1998 and references
therein; (5) Monaco et al. 2002; (6) Da Costa \& Mould 1988; (7) Han et al. 1997; (8) Martinez-Delgado \& Aparicio 1998;
(9) Geisler et al. 1999; (10) Mould, Kristian, \& Da Costa 1984; (11) Durrell et al. 1996a; (12) Puzia et al. 2001; (13) Kundu \& Whitmore 2001b;
(14) Durrell et al. 1996b}
\end{deluxetable}

\clearpage

\clearpage

\clearpage

\clearpage

\clearpage

\clearpage

\clearpage

\clearpage

\clearpage

\clearpage

\end{document}